\documentclass[12pt, journal, onecolumn]{IEEEtran}

\usepackage{setspace}
\usepackage{bm}

\ifCLASSINFOpdf
   \usepackage[pdftex]{graphicx}
\else
\fi

\usepackage[cmex10, tbtags]{amsmath}
\usepackage{amssymb}
\usepackage{dsfont}
\usepackage[tight,footnotesize]{subfigure}
\usepackage{color}

\newtheorem{lemma}{Lemma}[section]

\newtheorem{theorem}{Theorem}[section]
\newtheorem{corollary}{Corollary}[section]

\newtheorem{assumption}{Assumption}
\newtheorem{Remark}{\it Remark}[section]

\title{\begin{spacing}{1.3}Spatial Throughput Characterization in Cognitive Radio Networks with Threshold-Based Opportunistic Spectrum Access\footnote{X. Song, C. Yin, and D. Liu are with the Beijing Key Laboratory of Network System Architecture and Convergence, Beijing University of Posts and Telecommunications, Beijing, China, 100876. Emails: \{songxiaoshi, ccyin, dpliu\}@bupt.edu.cn.}
\footnote{R. Zhang is with the Department of Electrical and Computer Engineering, National University of Singapore, Singapore, 117576. Email: elezhang@nus.edu.sg.}\end{spacing}}
\author{Xiaoshi Song, Changchuan Yin, Danpu Liu, and Rui Zhang}

\begin{document}
\maketitle \thispagestyle{empty}
\vspace{-0.7in}
\begin{spacing}{1.5}
\begin{abstract}
This paper studies the opportunistic spectrum access (OSA) of the secondary users in a large-scale overlay cognitive radio (CR) network. Two threshold-based OSA schemes, namely the primary receiver assisted (PRA) protocol and the primary transmitter assisted (PTA) protocol, are investigated. Under the PRA/PTA protocol, a secondary transmitter (ST) is allowed to access the spectrum only when the maximum signal power of the received beacons/pilots sent from the active primary receivers/transmitters (PRs/PTs) is lower than a certain threshold. To measure the resulting transmission opportunity for the secondary users by the proposed OSA protocols, the concept of spatial opportunity, which is defined as the probability that an arbitrary location in the primary network is detected as a spatial spectrum hole, is introduced and then evaluated by applying tools from stochastic geometry.
Based on spatial opportunity, the coverage (non-outage transmission) performance in the overlay CR network is analyzed. With the obtained results of spatial opportunity and coverage probability, we finally characterize the spatial throughput, which is defined as the average spatial density of successful transmissions in the primary/secondary network, under the PRA and PTA protocols, respectively.
\end{abstract}

\end{spacing}
\vspace{-0.1in}

\section{Introduction}
Opportunistic spectrum access (OSA) \cite{DSA Survey:Zhao}, envisioned as a promising approach by utilizing cognitive radios (CRs) to improve the spectrum utilization efficiency, has attracted significant interests over the past few years. The basic idea of OSA is to enable the unlicensed secondary users to access the licensed spectrum by detecting and exploiting the spectrum holes available in the primary network. A spectrum hole, also referred to as underutilized position in the primary network, is defined as a multi-dimension (over time, frequency and space) region in which the transmission of a secondary transmitter (ST) introduces only limited interference at the active primary receivers (PRs)~\cite{Spectrum Hole:Anant Sahai}. Most of the existing literature has focused on the OSA design in time and/or frequency, where the STs utilize the idle time periods and/or frequency bands over the primary network to transmit \cite{CR Survey:Wang}--\cite{Review:Zeng}. In this paper, by applying tools from stochastic geometry \cite{Stochastic Geometry:Stoyan}--\cite{Stochastic Geometry: Martin}, we study the OSA design in space by exploiting the spatial spectrum holes in the primary network.

One challenging issue in the study of spatial OSA is how to measure the spatial spectrum availability in the primary network. To answer this question, in this paper, we introduce a new metric termed \textit{spatial opportunity}, which is defined as the probability that an arbitrary location in the geographical region of the primary network is detected as a spatial spectrum hole. Since spatial spectrum holes represent the underutilized positions that are expected to be reused by the secondary users, the spatial opportunity actually measures the spatial spectrum availability in the primary network and thereby quantifies the potential transmission opportunities for the secondary network.

Point process theory \cite{Point Process:Daley}, \cite{Book: Kingman} has been widely applied in the study of large-scale CR networks. In \cite{Generalized Results:Yin} and \cite{Transmission Capacity:Yin}, Yin \textit{et~al}. studied the transmission capacities of two overlaid mobile ad hoc networks (primary versus secondary) and analyzed their asymptotic capacity tradeoffs. In \cite{Multiple Antenna:Vaze}, Vaze evaluated the benefit of employing multiple antennas at secondary users by deriving the optimal spatial transmit/receive degrees of freedom for interference nulling/cancellation to maximize the scaling of the transmission capacity with respect to the number of antennas. In \cite{Cellular:Kaibin}, Huang \textit{et al}. studied the spectrum sharing between a cellular uplink versus a mobile ad hoc network and analyzed their performance tradeoff in terms of transmission capacity. In \cite{Spectrum-Sharing:Lee} and \cite{SIC:Lee}, J.~Lee \textit{et al}. investigated the spectrum sharing of multiple overlaid mobile ad hoc networks and characterized the effect of interference cancellation on the spectrum-sharing transmission capacity. It is worth noting that in the above prior works \cite{Generalized Results:Yin}--\cite{SIC:Lee}, since the spectrum access of the secondary user does not depend on the spatial realization of the primary users, the point processes formed by the active primary and secondary users are assumed to be independent.

In \cite{Exculsive Region:Martin}, C.~Lee \textit{et al}. investigated the aggregate interference and outage probabilities of the CR network in which the STs are allowed to transmit only if they are outside all the exclusion regions (guard zones) of the primary receivers/transmitters (PRs/PTs). Different from \cite{Generalized Results:Yin}--\cite{SIC:Lee}, in \cite{Exculsive Region:Martin}, due to the fact that the activation of secondary transmissions relies on the spatial realization of the primary network, the point processes formed by the active primary and secondary users become dependent, which gives rise to a new challenge in analyzing the outage performance. To tackle this challenge, the authors in \cite{Exculsive Region:Martin} characterized the conditional distribution of the locations of active PTs/STs given a typical primary/secondary receiver (PR/SR) at the origin via bounding and/or approximation techniques.

In this paper, we study a large-scale overlay CR network in which the STs are allowed to transmit only if they are detected to be in the spatial spectrum holes of the primary network. We consider time-slotted transmissions and two threshold-based OSA protocols described as follows.
\subsubsection{\textbf{Primary Receiver Assisted Protocol}}
In this protocol, to facilitate the OSA of the STs, each active PR is assumed to broadcast a unique beacon signal on a dedicated control channel \cite{Beacon:Hulbert}, \cite{Interference Aggregation:Ghasemi} at the beginning of each time slot. The STs are designed to monitor the beacon signals for $\text{spatial-spectrum-hole}$ detection. With the use of matched filtering\footnote{In this paper, for simplicity we assume perfect sensing, i.e., the noise effect on the matched filtering output is ignored.} \cite{CR Survey:Wang}, each ST is able to identify the beacons from different PRs and detect the maximum received beacon power. To simplify the analysis, as in \cite{Interference Aggregation:Ghasemi}, we assume that the primary control channel and data channel experience the same amount of path-loss and fading. As a result, due to channel reciprocity, the received beacon power can be used as a proxy for the STs to estimate their introduced interference perceived at each PR. To protect the primary transmissions, a predefined OSA threshold $N_{ra}$ is applied such that only the STs with the maximum received beacon power lower than the threshold are allowed to transmit. We call this threshold-based OSA protocol ``Primary Receiver Assisted (PRA)'' protocol.
\subsubsection{\textbf{Primary Transmitter Assisted Protocol}}
In this protocol, each active PT is assumed to transmit a unique pilot signal at the beginning of each time slot for coherent detection at the intended PR as in \cite{Generalized Results:Yin}--\cite{Exculsive Region:Martin}. The STs are designed to monitor the pilot signals from the active PTs for spatial-spectrum-hole detection. Similar to the case of the PRA protocol, with the use of matched filtering \cite{CR Survey:Wang}, each ST is able to identify the pilots from different PTs and detect the maximum received pilot power. To protect the primary transmissions, a predefined OSA threshold $N_{ta}$ is applied such that only the STs with the maximum received pilot power lower than the threshold are allowed to transmit. We call this threshold-based OSA protocol ``{Primary Transmitter Assisted} (PTA)'' protocol, to differ from PRA.

It is worth noting that under the proposed PRA or PTA protocol, due to the dependency of the secondary transmissions on the locations of active PRs/PTs, the point processes formed by the active primary and secondary users are in general {\it not independent}. As a result, how to characterize the dependency between the realizations of the active primary and secondary users under the proposed PRA or PTA protocol, which is different from that under the exclusion region based protocols proposed in \cite{Exculsive Region:Martin}, is the major challenge to be tackled in this paper. The main contributions of this paper are summarized as follows:
\begin{itemize}
  \item Two threshold-based OSA schemes, namely the PRA protocol and the PTA protocol, are proposed. To measure the resulting transmission opportunity for the STs by the proposed PRA or PTA protocol, the concept of spatial opportunity is introduced and evaluated by applying tools from stochastic geometry. Based on this result, the spatial distribution of the active STs under the PRA or PTA protocol is characterized.
  \item In both setups of the PRA and PTA protocols, given a typical PR/SR at the origin, the conditional distributions of the point processes formed by the active STs and/or PTs are derived based on the spatial opportunity characterization. It is worth noting that under the proposed PRA or PTA protocol, due to the threshold-based OSA, the point process formed by the active STs does not follow a homogeneous Poisson point process (HPPP) and thus a complete characterization of its spatial distribution is infeasible. As a result, the coverage (transmission non-outage) probability of the primary/secondary network, which depends on the Laplace transform of the aggregate interference from all active STs to the typical PR/SR at the origin, is difficult to be characterized exactly. To tackle this difficulty, in both setups of the PRA and PTA protocols, new approximations are made on the conditional distribution of the active STs. Based on such approximations, the coverage performance of the primary and secondary networks under the PRA or PTA protocol is characterized. Finally, with the results obtained from the analysis on spatial opportunity and coverage probability, we characterize the spatial throughput for both the primary and secondary networks under the PRA and PTA protocols, respectively.
\end{itemize}

{It is worth noting that the paper by Nguyen and Baccelli \cite{CSMA:Baccelli} is similar in spirit and scope to our work. However, our work differs from \cite{CSMA:Baccelli} in the following two main aspects. First, the protocol studied in our paper is different from that in \cite{CSMA:Baccelli}. Specifically, in \cite{CSMA:Baccelli} the authors considered the carrier sense multiple access (CSMA) based protocols, under which a primary user is allowed to access the spectrum only if it has the smallest back-off timer among its primary contenders while a secondary user is allowed to transmit only if it has no primary contenders and the corresponding back-off timer is the smallest among its secondary contenders. In contrast, we consider the Aloha type of protocol in this paper, where the primary users make independent decisions to access the spectrum with a given probability while the secondary users are allowed to transmit as long as they have no primary contenders detected. As a result, the distributions of active PTs and STs derived in our paper are considerably different from that in \cite{CSMA:Baccelli}. Second, our analysis is more general than that in \cite{CSMA:Baccelli}. Notice that in \cite{CSMA:Baccelli}, the primary and secondary users are assumed to have the same transmission parameters (e.g. transmit power, OSA threshold, node distance, SIR target). Then, by assigning proper virtual back-off timers, the authors in \cite{CSMA:Baccelli} regarded the primary and secondary networks as a single-layer ad hoc network and thereby were able to apply the results in \cite{Stochastic Geometry:Baccelli} (which are only applicable for the case of single-layer ad hoc networks) to characterize the distributions of active users (notice that in \cite{CSMA:Baccelli}, there was no fundamental difference between the conditional distributions of active primary and secondary users). In our paper, different from \cite{CSMA:Baccelli}, we model the primary and secondary users in two independent but interacting networks and derive the resulting distributions of active primary and secondary users even for the case when they have different transmission parameters.}

The remainder of this paper is organized as follows. The system model is described in Section~\uppercase\expandafter{\romannumeral 2}. The concept of spatial opportunity is introduced and characterized in Section~\uppercase\expandafter{\romannumeral 3}. The coverage performance of the primary and secondary networks is analyzed in Sections~\uppercase\expandafter{\romannumeral 4} and \uppercase\expandafter{\romannumeral 5}, respectively. Simulation results are presented in Section~\uppercase\expandafter{\romannumeral 6}. Finally, we conclude the paper in Section~\uppercase\expandafter{\romannumeral 7}.

Notations of selected symbols used in this paper are summarized in Table~\ref{Table:Notation}.

\begin{table}
\footnotesize
\begin{spacing}{1.3}
\caption{Symbol Notation}
\centering
\begin{tabular}{l l}
\hline
Symbol & Meaning \\
\hline

$\mu_0, \lambda_0$ & Density of initially deployed PTs/PRs, STs/SRs \\

$\mu_p$ & Density of active PTs/PRs \\

$\alpha$ & Path-loss exponent \\

$P_p, P_s$ & Transmission power of PTs, STs \\

$\theta_p, \theta_s$ & SIR target for the primary network, secondary network\\

{$N_{ra}, N_{ta}$} & {OSA threshold for STs under PRA, PTA protocol}\\

$Q_{ra}, Q_{ta}$ & Spatial opportunity for STs under PRA, PTA protocol \\

$\lambda_s^{ra}, \lambda_s^{ta}$ & Density of active STs under PRA, PTA protocol \\

${{\mathbf{R}}_p}, {{\mathbf{T}}_p}$ & Typical active PR, PT \\

${{\mathbf{R}}_s}, {{\mathbf{T}}_s}$ & Typical active SR, ST\\

$\Phi_{ra}^{{\bf{x}}}(u), \Phi_{ta}^{{\bf{x}}}(u)$ & Point process formed by the active STs on a circle of radius $u$ \\ &centered at location ${\bf{x}} \in \mathbb{R}^2$ under PRA, PTA protocol\\

$\Psi_{ra}^{{\bf{x}}}(u)$& Point process formed by the active PRs on a circle of radius $u$ \\ &centered at location ${\bf{x}} \in \mathbb{R}^2$ under PRA protocol\\

$\Upsilon_{ra}^{{\bf{x}}}(u), \Upsilon_{ta}^{{\bf{x}}}(u)$ & Point process formed by the active PTs on a circle of radius $u$ \\ &centered at location ${\bf{x}} \in \mathbb{R}^2$ under PRA, PTA protocol\\

$\lambda_{ra}^{{\bf{x}}}(u)$, $\lambda_{ta}^{{\bf{x}}}(u)$ & Density of $\Phi_{ra}^{{\bf{x}}}(u)$, $\Phi_{ta}^{{\bf{x}}}(u)$\\

$\psi_{ra}^{{\bf{x}}}(u)$ & Density of $\Psi_{ra}^{{\bf{x}}}(u)$\\

$\mu_{ra}^{{\bf{x}}}(u)$, $\mu_{ta}^{{\bf{x}}}(u)$ & Density of $\Upsilon_{ra}^{{\bf{x}}}(u), \Upsilon_{ta}^{{\bf{x}}}(u)$\\

$\tau_{p}^{ra}, \tau_{p}^{ta}$ & Coverage probability for the primary network under PRA, PTA protocol\\
$\tau_{s}^{ra}, \tau_{s}^{ta}$ & Coverage probability for the secondary network under PRA, PTA protocol\\
$C_{p}^{ra}, C_{p}^{ta}$ & Spatial throughput for the primary network under PRA, PTA protocol\\
$C_{s}^{ra}, C_{s}^{ta}$ & Spatial throughput for the secondary network under PRA, PTA protocol\\
\hline
\end{tabular}
\label{Table:Notation}
\end{spacing}
\end{table}

\section{Model and Metric}
\subsection{System Model}
We consider an overlay CR network in which two mobile ad hoc networks, namely the primary network and the secondary network, coexist and share the same spectrum on $\mathbb{R}^2$. The PTs are licensed users with a higher priority to access the spectrum, while the STs are allowed to transmit only if they are detected to be in the spatial holes of the primary network. The locations of the PTs and STs are assumed to follow two independent HPPPs with density $\mu_0$ and $\lambda_0$, respectively. For each PT, the intended PR is located at a distance of $d_p$ away in a random direction. Similarly, for each ST, the intended SR is located at a distance of $d_s$ away in a random direction. It should be noted that the PRs'/SRs' locations are not part of their respective transmitters' PPPs. Thus, the locations of the PRs and SRs follow two independent HPPPs with density $\mu_0$ and $\lambda_0$, respectively.

Assuming that time is slotted, and in each time slot the primary network employs an Aloha type of medium access control (MAC) protocol \cite{Spatial Aloha:Baccelli} such that the PTs make independent decisions to access the spectrum with probability $p_p$. Then, according to the coloring theorem \cite{Book: Kingman}, the locations of the active PTs/PRs follow a HPPP with density $\mu_p = \mu_0  p_p$.

For the secondary network, the PRA or PTA protocol is employed such that the STs are allowed to transmit only if they are detected to be in the resulting spatial spectrum holes of the primary network. It is worth noting that under the PRA or PTA protocol, unlike the position-independent thinning in the primary network, the access probabilities of the STs are position-dependent.
As such, the point process formed by the active STs under the PRA or PTA protocol does not follow a HPPP. In fact, under the PRA or PTA protocol, the access probability of each ST is a function of the realization of active PRs/PTs. Nevertheless, since the active PRs/PTs are homogeneous Poisson distributed, the access probabilities of all STs are identically distributed. Therefore, the randomly thinned point process formed by the active STs under the PRA or PTA protocol is stationary\footnote{A point process $\mathcal{N}$ is stationary if its characteristics are invariant under translation, i.e., the point processes $\mathcal{N} = \left\{{\bf{x}}_n\right\}$ and $\mathcal{N} = \left\{{\bf{x}}_n + {\bf{x}}\right\}$ have the same distribution for all ${\bf{x}} \in \mathbb{R}^2$ \cite{Stochastic Geometry:Stoyan}.} on $\mathbb{R}^2$.

The propagation channel is modeled as the combination of the small-scale Rayleigh fading and the large-scale path-loss given by
\begin{equation}\label{Channel Model}
g(d) = h d^{-\alpha},
\end{equation}
where $h$ denotes the exponentially distributed power coefficient with unit mean, $d$ denotes the propagation distance, and $\alpha$ denotes the path-loss exponent \cite{Threshold Scheduling:Weber}.

All the PTs are assumed to transmit the same power $P_p$. All the STs are assumed to transmit the same power $P_s$. In addition, all the PRs are assumed to use the same power $P_p$ for beacon transmissions. For the sake of simplicity, we ignore the thermal noise in the regime of interest and simply focus on the received signal-to-interference ratio (SIR) as in \text{\cite{Multiple Antenna:Vaze}--\cite{Exculsive Region:Martin}}. The SIR targets for the primary and secondary networks are denoted as $\theta_p$ and $\theta_s$, respectively.
\subsection{Performance Metric}
Three performance metrics are studied in this paper: the spatial opportunity, the coverage probability, and the spatial throughput, which are specified as follows.

\textbf{Spatial Opportunity:} The spatial opportunity in an overlay CR network, denoted by $Q$, is defined as the probability that a position $\bf{x} \in \mathbb{R}^2$ is detected as a spatial spectrum hole in the primary network with a given OSA policy (e.g. PRA or PTA).

\textbf{Coverage Probability:} The coverage probability, also known as the transmission non-outage probability, is defined as the probability that a (primary or secondary) receiver succeeds in decoding the received data packets from its corresponding (primary or secondary) transmitter. In particular, given the primary/secondary receiver SIR, denoted by $\mathrm{SIR}_p$ and $\mathrm{SIR}_s$, respectively, and the corresponding SIR targets, $\theta_p$ and $\theta_s$, the coverage probability in the primary/secondary network is defined as
\begin{equation}\label{Primary Coverage Probability Defination}
\tau_p = \Pr\left\{{\mathrm{SIR}}_p \geq \theta_p\right\},
\end{equation}
\begin{equation}\label{Secondary Coverage Probability Defination}
\tau_s = \Pr\left\{{\mathrm{SIR}}_s \geq \theta_s\right\}.
\end{equation}

\textbf{Spatial Throughput:} The spatial throughput of the primary/secondary network is the expected spatial density of successful primary/secondary transmissions, which are donated by $C_p$ and $C_s$, respectively, defined as
\begin{equation}\label{Primary Spatial Throughput Defination}
C_p = \mu_p \tau_p,
\end{equation}
\begin{equation}\label{Secondary Spatial Throughput Defination}
C_s = \lambda_0 Q \tau_s.
\end{equation}

\section{Spatial Opportunity}
Let $Q_{ra}$ and $Q_{ta}$ be the spatial opportunities of the overlay CR network under the proposed PRA and PTA protocols, respectively. Then, we characterize $Q_{ra}$ and $Q_{ta}$ in the following theorem.

\begin{theorem}\label{Theorem 1}
The spatial opportunity of an overlay CR network with the PRA/PTA protocol is given by
\begin{equation}\label{Spatial Opportunity for RA}
Q_{ra} = \exp\left\{- 2 \pi \mu_p \frac{\Gamma(\frac{2}{\alpha})(\frac{P_p}{N_{ra}})^{\frac{2}{\alpha}}}{\alpha} \right\},
\end{equation}
\begin{equation}\label{Spatial Opportunity for TA}
Q_{ta} = \exp\left\{- 2 \pi \mu_p \frac{\Gamma(\frac{2}{\alpha})(\frac{P_p}{N_{ta}})^{\frac{2}{\alpha}}}{\alpha} \right\},
\end{equation}
where {$N_{ra}$ and $N_{ta}$ denote the OSA threshold under the PRA and PTA protocols, respectively, and} $\Gamma(z)$ denotes the Gamma function with $z>0$, which is defined as
\[
\Gamma(z) = \int_0^{\infty} t^{z - 1} e^{-t} \textrm{d}t.
\]
\end{theorem}
\begin{IEEEproof}
See Appendix \ref{appendix 1}.
\end{IEEEproof}

The spatial opportunity quantifies the spatial spectrum availability in the primary network. The higher is the spatial opportunity, the more spatial locations in the primary network are being under-utilized. It is observed from (\ref{Spatial Opportunity for RA}) that the spatial opportunity is a function of the transmission parameters of the primary network.

On the other hand, since the STs are allowed to access the spectrum only if they are detected to be in the spatial spectrum holes of the primary network, the spatial opportunity actually measures the transmission opportunity for the secondary network. It is worth noting that $Q_{ra}$ and $Q_{ta}$ are obtained by averaging over all the possible realizations of the primary network. As a result, the spatial opportunity only quantifies the \emph{mean value} of the access probability of the STs in space. With this observation and by noting that $Q_{ra}$ and $Q_{ta}$ are position-independent, we have the following corollary.

\begin{corollary}\label{Corollary 1}
For an overlay CR network with the PRA/PTA protocol, the density of the point process formed by the active STs is given by $\lambda_s^{ra} = \lambda_0 Q_{ra}$ and $\lambda_s^{ta} = \lambda_0 Q_{ta}$, respectively.
\end{corollary}

From {Theorem~\ref{Theorem 1}}, it follows that if $N_{ra} = N_{ta}$, the spatial opportunities of the CR network with the PRA and PTA protocols are actually the same, i.e., ${Q_{ra} = Q_{ta}}$. However, since the detected spatial spectrum holes under the PRA and PTA protocols are distributed in a different manner, the coverage performance in the overlay CR network under these two setups are also different even with $N_{ra} = N_{ta}$. In the following two sections, we study the coverage probabilities of the primary and secondary networks, respectively, under the proposed PRA or PTA protocol.

\section{Coverage Probability in Primary Network}

\subsection{Conditional Distribution of Active STs}
\begin{figure*}[!h]
\centerline{\subfigure[]{\includegraphics[width
=2.6in]{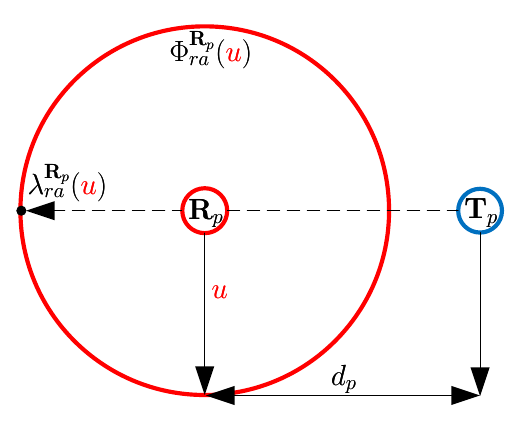}
\label{Distribution of RA}}
\hfil
\subfigure[]{\includegraphics[width=3.45in]{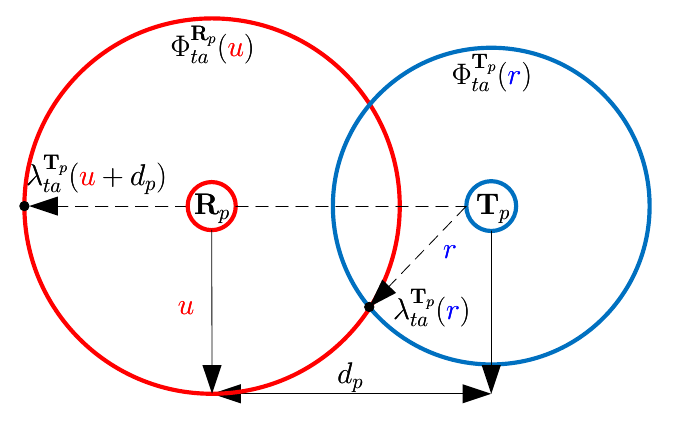}
\label{Distribution of TA}}}
\caption{Conditional distribution of active STs under (a) the PRA protocol; and (b) the PTA protocol.}
\end{figure*}

To analyze the coverage performance of the primary network, thanks to the stationarity of the point processes formed by the active primary and secondary users, we can focus on a typical PR at the origin denoted by ${{\mathbf{R}}_p}$ with its associated PT at a distance of $d_p$ away denoted by ${{\mathbf{T}}_p}$. Then, by Slivnyak's theorem \cite{Stochastic Geometry: Martin}, in both cases of the PRA and PTA protocols, the locations of the rest of the active PRs/PTs follow a HPPP with density $\mu_p$. For the secondary network, let $\Phi_{ra}^{{\mathbf{R}}_p}(u)$ be the point process formed by the active STs on a circle of radius $u$ centered at ${\mathbf{R}}_p$ under the PRA protocol as illustrated in Fig.~\ref{Distribution of RA}{\footnote{{More rigorously, we consider an annulus $\mathcal{A}_{ra}^{R_p}(u)$ bounded by two concentric circles centered at $R_p $ with radius of $u-\frac{\triangle u}{2}$ and $u +\frac{\triangle u}{2}$, respectively, and then define ${\Phi_{ra}^{R_p}(u)}$ as the point process formed by the active STs in $\mathcal{A}_{ra}^{R_p}(u)$ as $\triangle u \rightarrow 0$. In the following, with an abuse of notation, we simply denote ${\Phi_{ra}^{\bf{X}}(u)}$, ${\Phi_{ta}^{\bf{X}}(u)}$, ${\Psi_{ra}^{\bf{X}}(u)}$, ${\Upsilon_{ra}^{\bf{X}}(u)}$ and ${\Upsilon_{ta}^{\bf{X}}(u)}$ as the point processes formed by the active STs/PRs/PTs on a circle of radius $u$ centered at ${\bf{X}}$ under the PRA/PTA protocol to simplify the notation.}}}. In addition, let $\Phi_{ta}^{{\mathbf{T}}_p}(r)$ be the point process formed by the active STs on a circle of radius $r$ centered at ${\mathbf{T}}_p$ under the PTA protocol as illustrated in Fig.~\ref{Distribution of TA}. Then, we characterize the conditional distribution of the active STs under the PRA or PTA protocol in the following lemma.

\begin{lemma}\label{Lemma 1}
For an overlay CR network with the PRA/PTA protocol, conditioned on a typical PR at the origin, $\Phi_{ra}^{{\mathbf{R}}_p}(u)$ and $\Phi_{ta}^{{\mathbf{T}}_p}(r)$ are isotropic\footnote{A point process $\mathcal{N}$ is isotropic if its characteristics are invariant under rotation \cite{Stochastic Geometry:Stoyan}.} with respect to ${\mathbf{R}}_p$/${\mathbf{T}}_p$ with density $\lambda_{ra}^{{\mathbf{R}}_p}(u)$ and $\lambda_{ta}^{{\mathbf{T}}_p}(r)$, respectively, given by
\begin{equation}\label{the intensity with RA}
\lambda_{ra}^{{\mathbf{R}}_p}(u)
= \lambda_s^{ra}\left(1 - e^{-\frac{N_{ra} u^{\alpha}}{P_p}}\right),
\end{equation}
\begin{equation}\label{the intensity with TA}
\lambda_{ta}^{{\mathbf{T}}_p}(r) 
= \lambda_s^{ta}\left(1 - e^{-\frac{N_{ta} r^{\alpha}}{P_p}}\right).
\end{equation}
\end{lemma}
\begin{IEEEproof}
According to Theorem \ref{Theorem 1}, without conditioning on a typical PR at the origin, the spatial opportunity for a ST on a circle of radius $u$ or $r$ centered at ${{\mathbf{R}}_p}$ or ${{\mathbf{T}}_p}$ under the PRA or PTA protocol is given by $Q_{ra}$ or $Q_{ta}$. Conditioned on a typical PR at the origin, due to the newly introduced interference constraint at the typical PR, the spatial opportunity for a ST on the same circle centered at ${{\mathbf{R}}_p}$ or ${{\mathbf{T}}_p}$ under the PRA or PTA protocol reduces to $Q_{ra}\Pr\left(h \leq \frac{N_{ra} u^{\alpha}}{P_p}\right)$ or $Q_{ta}\Pr\left(h \leq \frac{N_{ta} r^{\alpha}}{P_p}\right)$, where $h$ denotes an exponentially distributed random variable with unit mean. Based on this result, it can be easily verified that $\Phi_{ra}^{{\mathbf{R}}_p}(u)$ or $\Phi_{ta}^{{\mathbf{T}}_p}(r)$ is isotropic around ${\mathbf{R}}_p$ or ${\mathbf{T}}_p$ with density $\lambda_{ra}^{{\mathbf{R}}_p}(u)$ or $\lambda_{ta}^{{\mathbf{T}}_p}(r)$ given by (\ref{the intensity with RA}) or (\ref{the intensity with TA}). This completes the proof of Lemma~\ref{Lemma 1}.~
\end{IEEEproof}

It is worth noting that under the PRA or PTA protocol, due to the threshold-based OSA, $\Phi_{ra}^{{\mathbf{R}}_p}(u)$ or $\Phi_{ta}^{{\mathbf{T}}_p}(r)$ does not follow a HPPP. Furthermore, since the higher order statistics of $\Phi_{ra}^{{\mathbf{R}}_p}(u)$ and $\Phi_{ta}^{{\mathbf{T}}_p}(r)$ are intractable, the coverage probability of the primary network under the PRA or PTA protocol, which depends on the Laplace transform of the aggregate interference from all active STs to the
typical PR at the origin, is difficult to be characterized exactly. To tackle this difficulty, similar to \cite{Exculsive Region:Martin}, {\cite{CSMA:Baccelli}}, \cite{Guard Zone:Andrews}, \cite{Energy:Zhang}, we make the following approximations on the conditional distribution of the active STs, which will be verified later by simulations in Section~\uppercase\expandafter{\romannumeral 6}.

\begin{assumption}
Under the PRA or PTA protocol, conditioned on a typical PR at the origin, $\Phi_{ra}^{{\mathbf{R}}_p}(u)$ or $\Phi_{ta}^{{\mathbf{T}}_p}(r)$ follows a HPPP with density $\lambda_{ra}^{{\mathbf{R}}_p}(u)$ or $\lambda_{ta}^{{\mathbf{T}}_p}(r)$ {and is assumed to be independent from the point process formed by the active PTs}.
\end{assumption}

Based on Assumption 1, we next characterize the coverage performance of the primary network under the PRA and PTA protocols in the following two subsections, respectively.

\subsection{Coverage Probability with PRA Protocol}
\begin{theorem}\label{Theorem 2}
For an overlay CR network with the PRA protocol, under Assumption 1, the coverage probability of the primary network is given by
\begin{equation}\label{Coverage Probability for Primary with RA Theorem}
\begin{split}
& \tau_p^{ra} = \exp\left\{-  \frac{2 \pi^2}{\alpha \sin\left(\frac{2 \pi}{\alpha}\right)}  \theta_p^{\frac{2}{\alpha}} d_p^2 \left(\mu_p + \lambda_s^{ra} \left(\frac{P_s}{P_p}\right)^{\frac{2}{\alpha}}\right) \right\} \times \exp\left\{\frac{2 \pi}{\alpha} \lambda_s^{ra} \left(\frac{P_p}{N_{ra}}\right)^{\frac{2}{\alpha}}\Gamma(\frac{2}{\alpha})\right\}\\
& \qquad\times \exp\left\{ - 2 \pi \lambda_s^{ra} \int_0^{\infty} \frac{\frac{P_p u^{\alpha}}{\theta_p P_s d_p^{\alpha}}}{1 + \frac{P_p u^{\alpha}}{\theta_p P_s d_p^{\alpha}}} \times
\frac{ e^{- \frac{\theta_p P_s N_{ra} d_p^{\alpha}}{P_p^2}}}{e^{\frac{N_{ra} u^{\alpha}}{P_p}}}u \textrm{d}u\right\}.
\end{split}
\end{equation}
\end{theorem}
\begin{IEEEproof}
See Appendix \ref{appendix 2}.
\end{IEEEproof}

\begin{Remark}
It is worth noting that under the PRA protocol, the point process formed by the active STs is isotropic around the typical PR at the origin. This is the key to that an exact characterization of the coverage probability of the primary network is obtained in Theorem \ref{Theorem 2}.
\end{Remark}

\begin{Remark}
It is also worth noting that as $N_{ra} \rightarrow \infty$ and $N_{ra} \rightarrow 0$, the coverage probability $\tau_p^{ra}$ is in accordance with that derived in \cite{Spectrum-Sharing:Lee} (where all the STs are active) and \cite{Multihop:Baccelli} (where none of the STs is active), respectively, as expected.
\end{Remark}

\begin{Remark}
With Theorem \ref{Theorem 2}, the spatial throughput of the primary network under the PRA protocol is given by
$C_p^{ra} = \mu_p \tau_p^{ra}$.
\end{Remark}
\subsection{Coverage Probability with PTA Protocol}
For an overlay CR network with the PTA protocol, let $\Phi_{ta}^{{\mathbf{R}}_p}(u)$ be the point process formed by the active STs on a circle of radius $u$ centered at the typical PR ${\mathbf{R}}_p$ as illustrated in Fig.~\ref{Distribution of TA}. Then, even with Assumption 1, $\Phi_{ta}^{{\mathbf{R}}_p}(u)$ follows a non-homogeneous PPP. Let $\lambda_{ta}^{{\mathbf{R}}_p}(u)$ be the average density of $\Phi_{ta}^{{\mathbf{R}}_p}(u)$. Then, we obtain the following two lemmas.
\begin{lemma}\label{Lemma 2}
For an overlay CR network with the PTA protocol, conditioned on a typical PR at the origin, an upper bound on $\lambda_{ta}^{{\mathbf{R}}_p}(u)$ is given by
\begin{align}\label{Bounds on the intensity with TA}
\lambda_{ta}^{{\mathbf{R}}_p}(u) \leq \lambda_s^{ta}\left(1 - e^{-\frac{N_{ta} \left(u + d_p\right)^{\alpha}}{P_p}}\right).
\end{align}
\end{lemma}
\begin{IEEEproof}
The proof immediately follows from Fig.~\ref{Distribution of TA} by observing that the highest density of $\Phi_{ta}^{{\mathbf{R}}_p}(u)$ is $\lambda_{ta}^{{\mathbf{T}}_p}(u + d_p)$.
\end{IEEEproof}
\begin{lemma}\label{Lemma 3}
For an overlay CR network with the PTA protocol, conditioned on a typical PR at the origin, the following inequality on $\lambda_{ta}^{{\mathbf{R}}_p}(u)$ holds:
\begin{equation}\label{lower Bound on the intensity with TA}
\int_0^{\infty}\frac{\lambda_{ta}^{{\mathbf{R}}_p}(u)}{1 + \frac{P_p u^{\alpha}}{\theta_p P_s d_p^{\alpha}}} u \textrm{d}u \geq \int_0^{\infty}\frac{\lambda_{ta}^{{\mathbf{T}}_p}(u)}{1 + \frac{P_p u^{\alpha}}{\theta_p P_s d_p^{\alpha}}} u \textrm{d}u.
\end{equation}
\end{lemma}
\begin{IEEEproof}
See Appendix \ref{appendix 3}.
\end{IEEEproof}

With Lemmas \ref{Lemma 2} and \ref{Lemma 3}, we are ready to characterize the coverage probability of the primary network under the PTA protocol, as given by the following theorem.
\begin{theorem}\label{Theorem 3}
For an overlay CR network with the PTA protocol, under Assumption 1, the coverage probability of the primary network is upper-bounded and lower-bounded, respectively, by
\begin{equation}\label{Upper Bound on the Coverage Probability for Primary with TA}
\begin{split}
\!\!\tau_p^{ta}  \leq &\exp\left\{-  \frac{2 \pi^2}{\alpha \sin\left(\frac{2 \pi}{\alpha}\right)}  \theta_p^{\frac{2}{\alpha}} d_p^2 \left(\mu_p + \lambda_s^{ta} \left(\frac{P_s}{P_p}\right)^{\frac{2}{\alpha}}\right) \right\}\\
&\times \exp\left\{ 2 \pi \lambda_s^{ta} \int_0^{\infty} \left(\frac{e^{ - \frac{N_{ta} u^{\alpha}}{P_p}}}{1 + \frac{P_p u^{\alpha}}{\theta_p P_s d_p^{\alpha}}} \right)u \textrm{d}u\right\},
\end{split}
\end{equation}
\begin{equation}\label{Lower Bound on the Coverage Probability for Primary with TA}
\begin{split}
\!\!\tau_p^{ta}  \geq &\exp\left\{-  \frac{2 \pi^2}{\alpha \sin\left(\frac{2 \pi}{\alpha}\right)}  \theta_p^{\frac{2}{\alpha}} d_p^2 \left(\mu_p + \lambda_s^{ta} \left(\frac{P_s}{P_p}\right)^{\frac{2}{\alpha}}\right) \right\}\\
&\times \exp\left\{ 2 \pi \lambda_s^{ta} \int_0^{\infty} \left(\frac{e^{ - \frac{N_{ta} (u + d_p)^{\alpha}}{P_p}}}{1 + \frac{P_p u^{\alpha}}{\theta_p P_s d_p^{\alpha}}} \right)u \textrm{d}u\right\}.
\end{split}
\end{equation}
\end{theorem}
\begin{IEEEproof}
See Appendix \ref{appendix 4}.
\end{IEEEproof}

\begin{Remark}
It is worth noting that under the PTA protocol, the point process formed by the active STs is isotropic around the typical PT instead of the corresponding typical PR at the origin. This results in that, unlike the case of PRA protocol, only the upper and lower bounds on the coverage probability of the primary network are obtained for PTA protocol in Theorem \ref{Theorem 3}.
\end{Remark}

\begin{Remark}{
Let $\triangle \tau_p^{ta}$ be the gap between the upper and lower bounds of $\tau_p^{ta}$. Then, from (\ref{Upper Bound on the Coverage Probability for Primary with TA}) and (\ref{Lower Bound on the Coverage Probability for Primary with TA}), it can be easily verified that $\mathop {\lim }\limits_{{d_p} \to 0} \triangle \tau_p^{ta} = 0$. Intuitively, as $d_p \to 0$, the density of active STs around the typical PR becomes the same as that around the typical PT. This is the reason why the upper and lower bounds of $\tau_p^{ta}$ converge as $d_p \to 0$ (Please see the proof for Theorem 4.2 in Appendix D for details.). On the other hand, due to the fact that the first term in the expressions of both the upper and lower bounds, i.e., $\exp\left\{-  \frac{2 \pi^2}{\alpha \sin\left(\frac{2 \pi}{\alpha}\right)}  \theta_p^{\frac{2}{\alpha}} d_p^2 \left(\mu_p + \lambda_s^{ta} \left(\frac{P_s}{P_p}\right)^{\frac{2}{\alpha}}\right) \right\}$, goes to $0$ as $d_p \to \infty$, we have $\mathop {\lim }\limits_{{d_p} \to \infty} \triangle \tau_p^{ta} = 0$. Similar results can also be obtained with respect to $N_{ta}$, i.e., $\mathop {\lim }\limits_{{N_{ta}} \to 0} \triangle \tau_p^{ta} = 0$ and $\mathop {\lim }\limits_{{N_{ta}} \to \infty} \triangle \tau_p^{ta} = 0$. It is worth noting that due to the complex integrals in (\ref{Upper Bound on the Coverage Probability for Primary with TA}) and (\ref{Lower Bound on the Coverage Probability for Primary with TA}), it is difficult to find the maximum value of $\triangle \tau_p^{ta}$ with respect to $d_{p}$ or $N_{ta}$. However, with (\ref{Upper Bound on the Coverage Probability for Primary with TA}) and (\ref{Lower Bound on the Coverage Probability for Primary with TA}), the maximum value of $\triangle \tau_p^{ta}$ can be numerically obtained. It is also worth noting that, in general, $\triangle \tau_p^{ta}$ depends on other parameters as well (e.g. $\lambda_0$, as can be observed in Fig.~\ref{Primary Coverage Probability versus N_th}).}
\end{Remark}

\begin{Remark}
Based on Theorem \ref{Theorem 3}, we thereby characterize the upper and lower bounds on the spatial throughput $C_p^{ta} = \mu_p \tau_p^{ta}$ of the primary network under the PTA protocol.
\end{Remark}

\section{Coverage Probability in Secondary Network}
\subsection{Conditional Distributions of Active PTs and STs}
To analyze the coverage performance of the secondary network, we focus on a typical SR at the origin denoted by ${\mathbf{R}}_s$ with its associated ST at a distance of $d_s$ away denoted by ${\mathbf{T}}_s$. Let ${\Psi}_{ra}^{{\mathbf{T}}_s}(r)$ and ${\Upsilon}_{ta}^{{\mathbf{T}}_s}(r)$ be the point processes formed by the active PRs and PTs, respectively, on a circle of radius $r$ centered at ${\mathbf{T}}_s$ under the PRA/PTA protocol. Then, the conditional distribution of the active PRs/PTs under the PRA/PTA protocol is characterized as follows.
\begin{lemma}\label{Lemma 4}
For an overlay CR network with the PRA/PTA protocol, conditioned on a typical SR at the origin, ${\Psi}_{ra}^{{\mathbf{T}}_s}(r)$ and ${\Upsilon}_{ta}^{{\mathbf{T}}_s}(r)$ are HPPPs with their respective densities given by
\begin{equation}\label{Secondary PR the intensity with RA}
{\psi}_{ra}^{{\mathbf{T}}_s}(r) = \mu_p \left(1 - e^{-\frac{N_{ra} r^{\alpha}}{P_p}}\right),
\end{equation}
\begin{equation}\label{Secondary PR the intensity with TA}
{\mu}_{ta}^{{\mathbf{T}}_s}(r) = \mu_p \left(1 - e^{-\frac{N_{ta} r^{\alpha}}{P_p}}\right).
\end{equation}
\end{lemma}
\begin{IEEEproof}
Under the PRA or PTA protocol, conditioned on a typical SR at the origin, the probability that a PR or PT on a circle of radius $r$ centered at ${\mathbf{T}}_s$ is active is given by $\Pr\left(h \leq \frac{N_{ra} r^{\alpha}}{P_p}\right)$ or $\Pr\left(h \leq \frac{N_{ta} r^{\alpha}}{P_p}\right)$, where $h$ denotes an exponentially distributed random variable with unit mean. Then, according to the coloring theorem \cite{Book: Kingman}, it can be easily verified that ${\Psi}_{ra}^{{\mathbf{T}}_s}(r)$ or ${\Upsilon}_{ta}^{{\mathbf{T}}_s}(r)$ follows a HPPP with density ${\psi}_{ra}^{{\mathbf{T}}_s}(r)$ or ${\mu}_{ta}^{{\mathbf{T}}_s}(r)$ as given by (\ref{Secondary PR the intensity with RA}) or (\ref{Secondary PR the intensity with TA}). This completes the proof of Lemma~\ref{Lemma 4}~.~
\end{IEEEproof}

Let ${\Upsilon}_{ra}^{{\mathbf{T}}_s}(r)$ be the point process formed by the active PTs on a circle of radius $r$ centered at ${\mathbf{T}}_s$ under the PRA protocol. Then, based on Lemma \ref{Lemma 4}, we characterize the conditional distribution of the active PTs under the PRA protocol in the following lamma.
\begin{lemma}\label{Lemma 6}
For an overlay CR network with the PRA protocol, conditioned on a typical SR at the origin, ${\Upsilon}_{ra}^{{\mathbf{T}}_s}(r)$ follows a HPPP with density ${\mu}_{ra}^{{\mathbf{T}}_s}(r)$, which is upper-bounded by
\begin{equation}\label{Secondary PR the intensity with RA 3}
\begin{split}
{\mu}_{ra}^{{\mathbf{T}}_s}(r) &\leq \mu_p \left(1 - e^{-\frac{N_{ra} (r + d_p)^{\alpha}}{P_p}}\right).
\end{split}
\end{equation}
\end{lemma}
\begin{IEEEproof}
The conditional distribution of the active PTs is related to that of their corresponding active PRs located at a distance of $d_p$ away in random directions. From Lemma~\ref{Lemma 4}, it thus follows that ${\Upsilon}_{ra}^{{\mathbf{T}}_s}(r)$ is a HPPP with density ${\mu}_{ra}^{{\mathbf{T}}_s}(r)$, which is (in the worst case) upper-bounded by ${\psi}_{ra}^{{\mathbf{T}}_s}(r + d_p)$.
This completes the proof of Lemma \ref{Lemma 6}.
\end{IEEEproof}

\begin{figure*}[!h]
\centerline{\subfigure[]{\includegraphics[width
=3.3in]{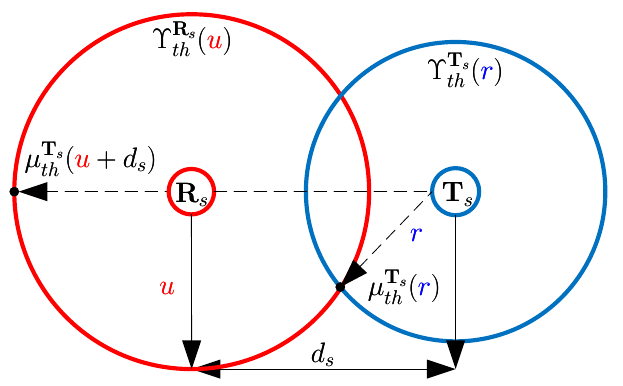}
\label{Conditional Distribution of PTs}}
\hfil
\subfigure[]{\includegraphics[width=2.6in]{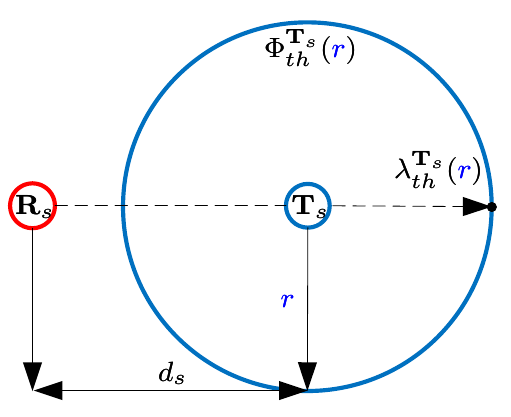}
\label{Conditional Distribution of STs}}}
\caption{Illustration of the conditional distributions for (a) active PTs; and (b) active STs, under the PRA or PTA protocol, where $th$ stands for $ra$ or $ta$.}
\end{figure*}

The conditional distribution of the active PTs under the PRA or PTA protocol is illustrated in Fig.~\ref{Conditional Distribution of PTs}.

For the secondary network, let ${\Phi}_{ra}^{{\mathbf{T}}_s}(r)$ or ${\Phi}_{ta}^{{\mathbf{T}}_s}(r)$ be the point process formed by the active STs on a circle of radius $r$ centered at ${\mathbf{T}}_s$ under the PRA or PTA protocol as illustrated in Fig.~\ref{Conditional Distribution of STs}. Then, based on Lemma \ref{Lemma 4}, the conditional distribution of the active STs under the PRA or PTA protocol is characterized as follows.
\begin{lemma}\label{Lemma 5}
For an overlay CR network with the PRA or PTA protocol, conditioned on a typical SR at the origin, ${\Phi}_{ra}^{{\mathbf{T}}_s}(r)$ and ${\Phi}_{ta}^{{\mathbf{T}}_s}(r)$ are isotropic around ${\mathbf{T}}_s$, and their densities, denoted by ${\lambda^{{\mathbf{T}}_s}_{ra}}({r})$ and ${\lambda^{{\mathbf{T}}_s}_{ta}}({r})$, respectively, are bounded by
\begin{equation}\label{Secondary Coverage Secondary Density with RA}
\lambda_s^{ra} \leq {\lambda^{{\mathbf{T}}_s}_{ra}}({r}) \leq \lambda_s^{ra}\beta_{ra},
\end{equation}
\begin{equation}\label{Secondary Coverage Secondary Density with TA}
\lambda_s^{ta} \leq {\lambda^{{\mathbf{T}}_s}_{ta}}({r}) \leq \lambda_s^{ta}\beta_{ta},
\end{equation}
where
\begin{equation}\label{Secondary Saptial Opportunity with RA 1}
\beta_{ra} = \exp\left\{\pi \mu_p {\Gamma(\frac{2 + \alpha}{\alpha})(\frac{P_p}{2 N_{ra}})^{\frac{2}{\alpha}}} \right\},
\end{equation}
\begin{equation}\label{Secondary Saptial Opportunity with TA 1}
\beta_{ta} = \exp\left\{\pi \mu_p {\Gamma(\frac{2 + \alpha}{\alpha})(\frac{P_p}{2 N_{ta}})^{\frac{2}{\alpha}}} \right\}.
\end{equation}
\end{lemma}
\begin{IEEEproof}
See Appendix \ref{appendix 5}.
\end{IEEEproof}

It is worth noting that under the PRA or PTA protocol, similar to the primary network case, ${\Phi}_{ra}^{{\mathbf{T}}_s}(r)$ or ${\Phi}_{ta}^{{\mathbf{T}}_s}(r)$ does not follow a HPPP. As a result, with only the first-order moment measures (average densities) of ${\Phi}_{ra}^{{\mathbf{T}}_s}(r)$ and ${\Phi}_{ta}^{{\mathbf{T}}_s}(r)$ being obtained, the coverage probability of the secondary network under the PRA or PTA protocol is difficult to be characterized exactly. To tackle this difficulty, we make the following approximation on the conditional distribution of the active STs, which will be verified later by simulations in Section~\uppercase\expandafter{\romannumeral 6}.
\begin{assumption}
Under the PRA or PTA protocol, conditioned on a typical SR at the origin, $\Phi_{ra}^{{\mathbf{T}}_s}(r)$ or $\Phi_{ta}^{{\mathbf{T}}_s}(r)$ follows a HPPP with density $\lambda_{ra}^{{\mathbf{T}}_s}(r)$ or $\lambda_{ta}^{{\mathbf{T}}_s}(r)$ {and is assumed to be independent from the point process formed by the active PTs}.
\end{assumption}

Based on Assumption 2, we characterize the coverage performance of the secondary network under the PRA and PTA protocols in the following two subsections, respectively.
\subsection{Coverage Probability with PRA Protocol}
Under the PRA protocol, let ${\Upsilon}_{ra}^{{\mathbf{R}}_s}(u)$ be the point process formed by the active PTs on a circle of radius $u$ centered at ${\mathbf{R}}_s$ as illustrated in Fig.~\ref{Conditional Distribution of PTs}. From Lemma \ref{Lemma 6}, it thus follows that in general ${\Upsilon}_{ra}^{{\mathbf{R}}_s}(u)$ is a non-homogeneous PPP. Let ${\mu}_{ra}^{{\mathbf{R}}_s}(u)$ be the average density of ${\Upsilon}_{ra}^{{\mathbf{R}}_s}(u)$. Then, we obtain the following lemma.
\begin{lemma}\label{Lemma 7}
Under the PRA protocol, conditioned on a typical SR at the origin, an upper bound on ${\mu}_{ra}^{{\mathbf{R}}_s}(u)$ is given by
\vspace{-0.2in}
\begin{equation}\label{Secondary PR the intensity with RA 6}
\begin{split}
{\mu}_{ra}^{{\mathbf{R}}_s}(u) &\leq \mu_p \left(1 - e^{-\frac{N_{ra} (u + d_p + d_s)^{\alpha}}{P_p}}\right).
\end{split}
\end{equation}
\end{lemma}
\begin{IEEEproof}
Similar to the proof of Lemma \ref{Lemma 2}, (\ref{Secondary PR the intensity with RA 6}) is obtained directly from Lemma \ref{Lemma 6}.
\end{IEEEproof}

Now, we are ready to evaluate the coverage probability of the secondary network
under the PRA protocol, as given by the following theorem.
\begin{theorem}\label{Theorem 4}
For an overlay CR network with the PRA protocol, under Assumption 2, the coverage probability of the secondary network is lower-bounded by
\begin{equation}\label{Lower Bound Coverage Probability for Secondary with RA}
\begin{split}
\tau_s^{ra} &\geq \exp\left\{ - \frac{2 \pi^2}{\alpha \sin\left(\frac{2 \pi}{\alpha}\right)}  \theta_s^{\frac{2}{\alpha}} d_s^2 \left( \mu_p \left(\frac{P_p}{P_s}\right)^{\frac{2}{\alpha}} + \lambda_s^{ra}\beta_{ra}\right)\!\right\}\\
&\times \exp\left\{ 2 \pi \mu_p \int_0^{\infty} \left(\frac{e^{ - \frac{N_{ra} (u + d_p + d_s)^{\alpha}}{P_p}}}{1 + \frac{P_s u^{\alpha}}{\theta_s P_p d_s^{\alpha}}} \right)u \textrm{d}u\right\}.
\end{split}
\end{equation}
\end{theorem}
\begin{IEEEproof}
With Lemmas \ref{Lemma 5} and \ref{Lemma 7}, (\ref{Lower Bound Coverage Probability for Secondary with RA}) is readily obtained by applying a similar approach as for the proof of {Theorem~\ref{Theorem 3}}.~
\end{IEEEproof}


\begin{Remark}
With Theorems \ref{Theorem 1} and \ref{Theorem 4}, we thus establish a lower bound on the spatial throughput  $C_s^{ra} = \lambda_0 Q_{ra} \tau_s^{ra}$ of the secondary network under the PRA protocol.
\end{Remark}
\subsection{Coverage Probability with PTA Protocol}
Under the PTA protocol, let ${\Upsilon}_{ta}^{{\mathbf{R}}_s}(u)$ be the point process formed by the active PTs on a circle of radius $u$ centered at ${\mathbf{R}}_s$ as illustrated in Fig.~\ref{Conditional Distribution of PTs}. From Lemma \ref{Lemma 4}, it is known that ${\Upsilon}_{ta}^{{\mathbf{R}}_s}(u)$ follows a non-homogeneous PPP. Let $\mu_{ta}^{{\mathbf{R}}_s}(u)$ be the average density of ${\Upsilon}_{ta}^{{\mathbf{R}}_s}(u)$. Then, we obtain the following two lemmas.
\begin{lemma}\label{Lemma 8}
Under the PTA protocol, conditioned on a typical SR at the origin, an upper bound on ${\mu}_{ta}^{{\mathbf{R}}_s}(u)$ is given by
\vspace{-0.2in}
\begin{equation}\label{Secondary PR the intensity with TA 6}
\begin{split}
{\mu}_{ta}^{{\mathbf{R}}_s}(u) &\leq \mu_p \left(1 - e^{-\frac{N_{ta} (u + d_s)^{\alpha}}{P_p}}\right).
\end{split}
\end{equation}
\end{lemma}
\begin{IEEEproof}
The proof is similar to that for Lemma \ref{Lemma 2} and thus is omitted for brevity.
\end{IEEEproof}

\begin{lemma}\label{Lemma 9}
Under the PTA protocol, conditioned on a typical SR at the origin, the following inequality on ${\mu}_{ta}^{{\mathbf{R}}_s}(u)$ holds:
\begin{equation}\label{lower Bound on the intensity with TA 1}
\int_0^{\infty}\frac{\mu_{ta}^{{\mathbf{R}}_s}(u)}{1 + \frac{P_p u^{\alpha}}{\theta_p P_s d_p^{\alpha}}} u \textrm{d}u \geq  \int_0^{\infty}\frac{\mu_{ta}^{{\mathbf{T}}_s}(u)}{1 + \frac{P_p u^{\alpha}}{\theta_p P_s d_p^{\alpha}}} u \textrm{d}u.
\end{equation}
\end{lemma}
\begin{IEEEproof}
The proof is similar to that for Lemma \ref{Lemma 3} and thus is omitted for brevity.
\end{IEEEproof}

With Lemmas \ref{Lemma 8} and \ref{Lemma 9}, we are ready to evaluate the coverage probability of the secondary network under the PTA protocol, as given by the following theorem.
\begin{theorem}\label{Theorem 5}
For an overlay CR network with the PTA protocol, under Assumption 2, the coverage probability of the secondary network is upper-bounded and lower-bounded, respectively, by
\begin{equation}\label{Upper Bound on the Coverage Probability for Secondary with TA}
\begin{split}
\!\!\tau_s^{ta}  \leq &\exp\left\{-  \frac{2 \pi^2}{\alpha \sin\left(\frac{2 \pi}{\alpha}\right)}  \theta_s^{\frac{2}{\alpha}} d_s^2 \left(\mu_p \left(\frac{P_p}{P_s}\right)^{\frac{2}{\alpha}} + \lambda_s^{ta} \right)\right\}\\
&\times \exp\left\{ 2 \pi \mu_p  \int_0^{\infty} \left(\frac{e^{ - \frac{N_{ta} u^{\alpha}}{P_p}}}{1 + \frac{P_s u^{\alpha}}{\theta_s P_p d_s^{\alpha}}} \right)u \textrm{d}u\right\},
\end{split}
\end{equation}
\begin{equation}\label{Lower Bound on the Coverage Probability for Secondary with TA}
\begin{split}
\!\!\!\!\!\tau_s^{ta}  \geq &\exp\left\{-  \frac{2 \pi^2}{\alpha \sin\left(\frac{2 \pi}{\alpha}\right)}  \theta_s^{\frac{2}{\alpha}} d_s^2 \left(\mu_p \left(\frac{P_p}{P_s}\right)^{\frac{2}{\alpha}} + \lambda_s^{ta} \beta_{ta} \right)\right\}\\
&\times \exp\left\{ 2 \pi \mu_p  \int_0^{\infty} \left(\frac{e^{ - \frac{N_{ta} (u + d_s)^{\alpha}}{P_p}}}{1 + \frac{P_s u^{\alpha}}{\theta_s P_p d_s^{\alpha}}} \right)u \textrm{d}u\right\}.
\end{split}
\end{equation}
\end{theorem}
\begin{IEEEproof}
Based on Lemmas \ref{Lemma 5}, \ref{Lemma 8} and \ref{Lemma 9}, and by applying a similar approach as in the proof of {Theorem~\ref{Theorem 3}}, (\ref{Upper Bound on the Coverage Probability for Secondary with TA}) and (\ref{Lower Bound on the Coverage Probability for Secondary with TA}) are thus obtained.~
\end{IEEEproof}

\begin{Remark}
It is worth noting that in the case of PTA protocol, the conditional distribution of active PTs can be exactly characterized. For this reason, both the upper and lower bounds are obtained for the coverage probability of the secondary network as shown in Theorem \ref{Theorem 5}.
\end{Remark}

\begin{Remark}
It is also worth noting that as $N_{ta} \rightarrow 0$, according to (\ref{Upper Bound on the Coverage Probability for Secondary with TA}) and (\ref{Lower Bound on the Coverage Probability for Secondary with TA}), we have
\[
 \exp\left\{-  \frac{2 \pi^2}{\alpha \sin\left(\frac{2 \pi}{\alpha}\right)}  \theta_s^{\frac{2}{\alpha}} d_s^2 \lambda_s^{ta} \beta_{ta}\right\} \leq \tau_s^{ta} \leq \exp\left\{-  \frac{2 \pi^2}{\alpha \sin \left(\frac{2 \pi}{\alpha}\right)}  \theta_s^{\frac{2}{\alpha}} d_s^2 \lambda_s^{ta}\right\}.
\]
Therefore, as $N_{ta} \rightarrow 0$, the coverage performance of the secondary network is solely determined by the secondary transmissions. An intuitive explanation of the above observation is that, when $N_{ta}$ is small, the active PTs are in general very far away from the typical SR at the origin and thus the nearby active STs dominate the coverage performance of the secondary network.
\end{Remark}

\begin{Remark}
With Theorems \ref{Theorem 1} and \ref{Theorem 5}, the spatial throughput $C_s^{ta} = \lambda_0 Q_{ta} \tau_s^{ta}$ of the secondary network under the PTA protocol is thereby characterized.
\end{Remark}

\section{Numerical Results}
In this section, we present simulation results on the performance of the PRA and PTA protocols to validate our analytical results. Throughout this section, unless specified otherwise, we set $\mu_p = 0.01$, $P_p = 5$, $P_s = 2$, $d_p = d_s = 1$, $\theta_p = \theta_s = 3$, and $\alpha = 4$.
\subsection{Spatial Opportunity}
\begin{figure}[!h]
\centering
\includegraphics[width=3.5in]{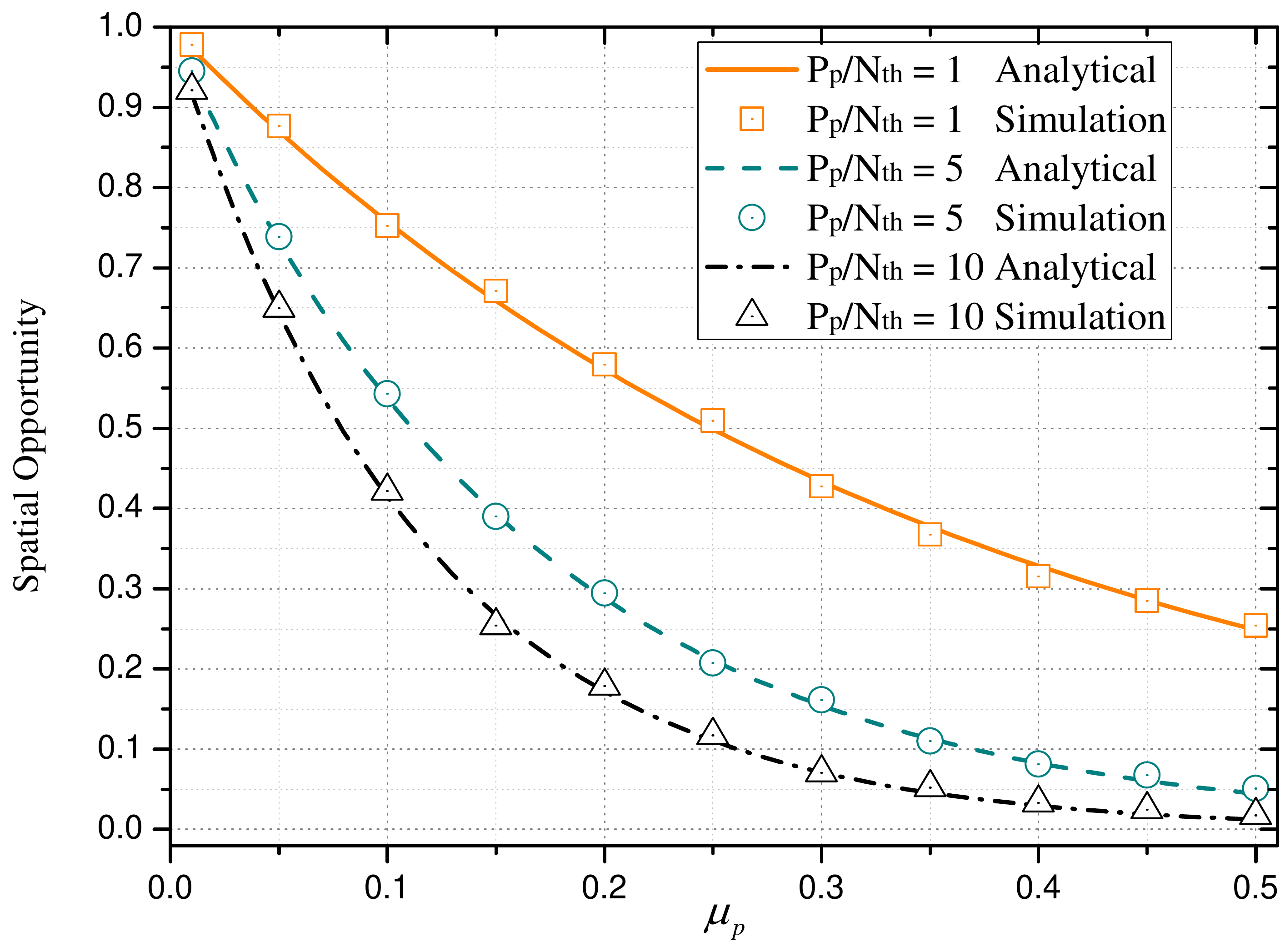}
\caption{Spatial opportunity in an overlay CR network versus the active primary user density $\mu_p$ under the PRA/PTA protocol, with ${N_{ra} = N_{ta} = N_{th}}$. } \label{Spatial Opportunity Plot}
\end{figure}

Fig.~\ref{Spatial Opportunity Plot} shows the analytical and simulated results on the spatial opportunity $Q_{ra}$ by the PRA protocol or $Q_{ta}$ by the PTA protocol versus the active primary user density $\mu_p$ when ${P_p}/{N_{th}} = 1, 5, \text{and } 10$, respectively, where we set $N_{ra} = N_{ta} = N_{th}$. It is observed that the spatial opportunity in an overlay CR network with threshold-based OSA is a decreasing function of $\mu_p$ as well as the ratio $P_p/N_{th}$, which are expected according to {Theorem~\ref{Theorem 1}}. It is also observed that the simulation results fit closely to our analytical results.
\subsection{Coverage Probability}
\begin{figure*}[!h]
\centerline{\subfigure[$\lambda_0 = 0.01$]{\includegraphics[width
=3.5in]{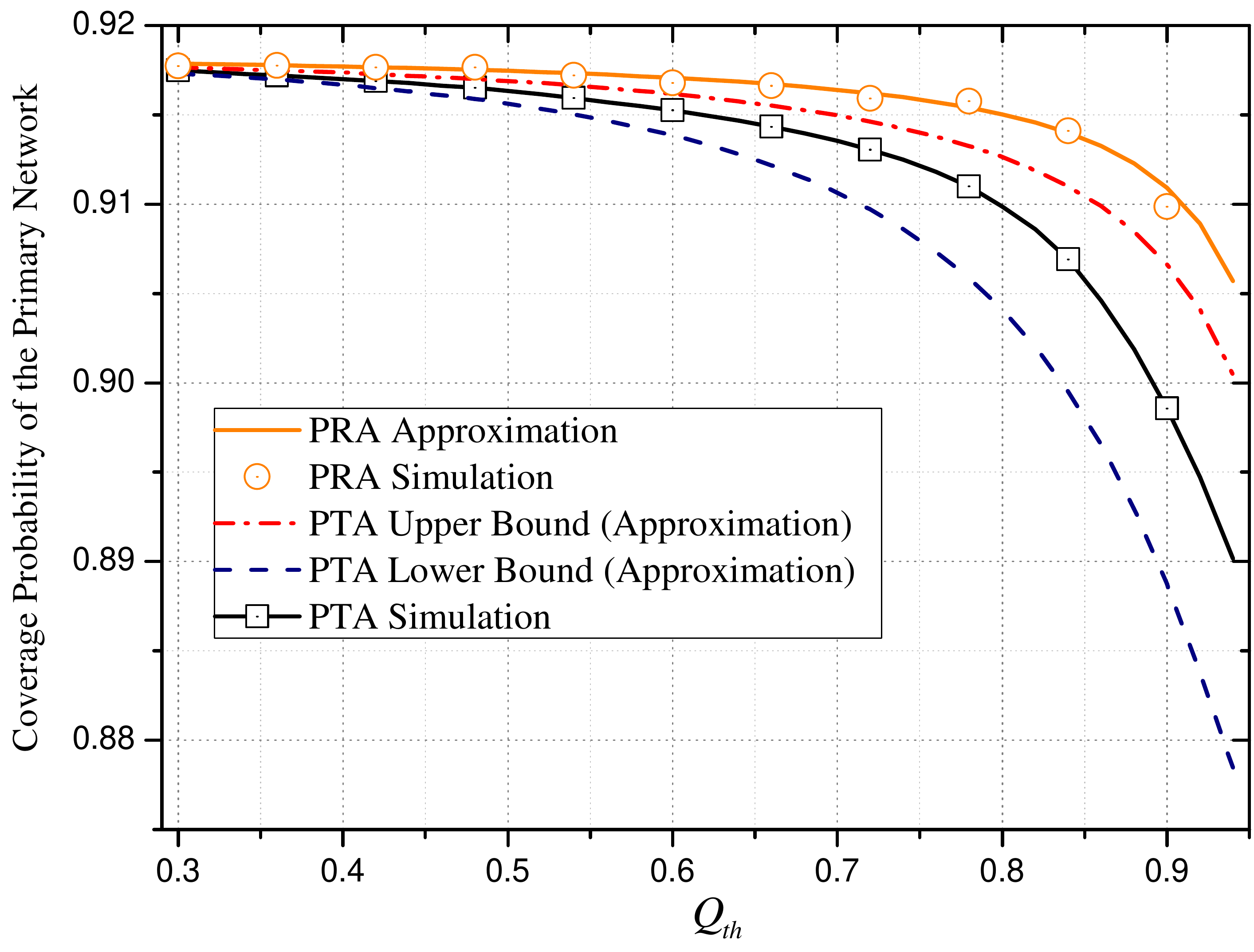}}
\hfil
\subfigure[${\lambda_0 = 0.1}$]{\includegraphics[width=3.5in]{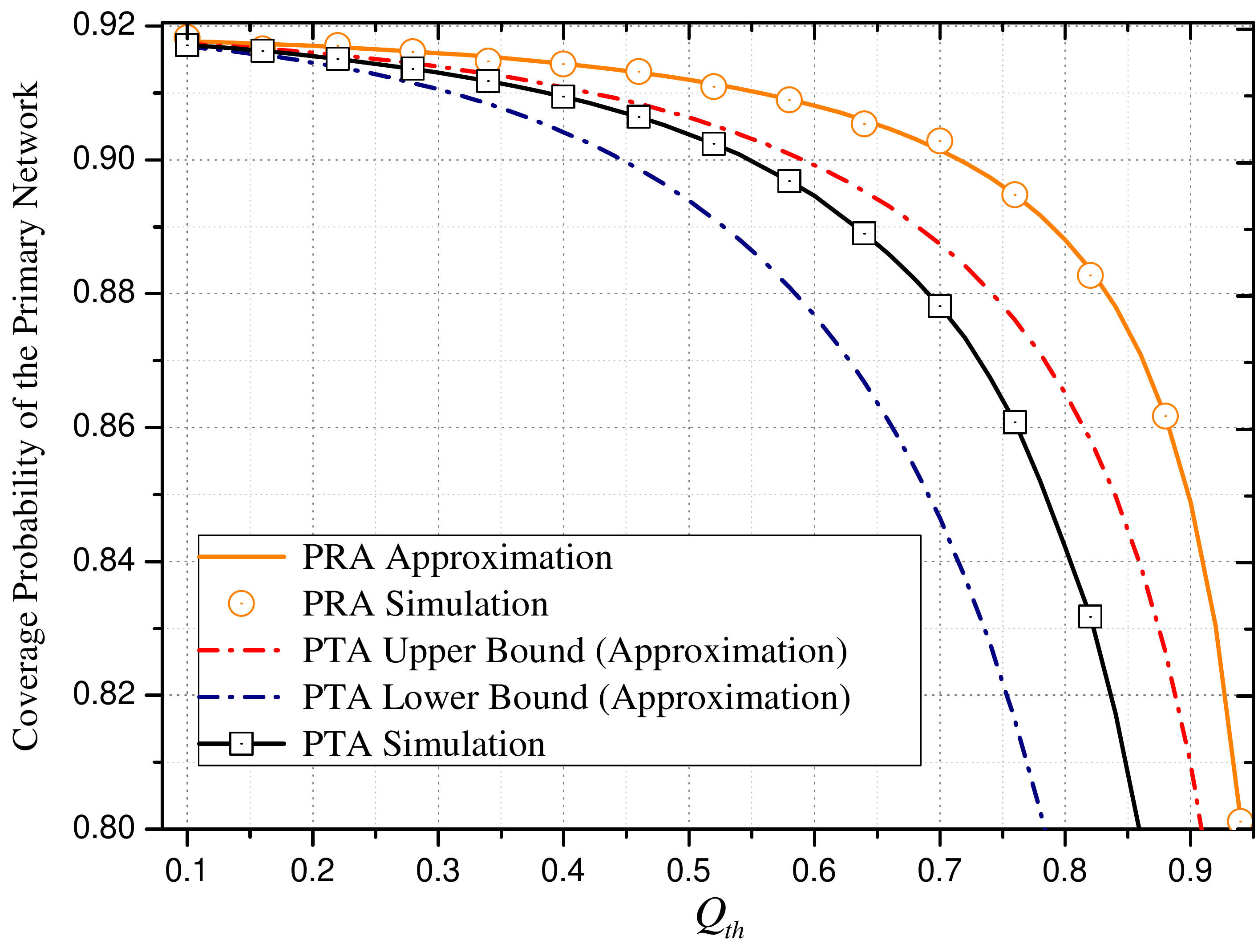}}}
\caption{Coverage probability of the primary network versus spatial opportunity $Q_{th}$ under the PRA/PTA protocol, with ${Q_{ra} = Q_{ta} = Q_{th}}$.}
\label{Primary Coverage Probability versus N_th}
\end{figure*}

In Fig.~\ref{Primary Coverage Probability versus N_th}, we compare the analytical and simulated results on the coverage probability of the primary network under the PRA and PTA protocols, respectively. Several observations are in order. First, the approximated coverage probability of the primary network under the PRA protocol derived in {Theorem~\ref{Theorem 2}} under Assumption 1 is quite accurate. An intuitive explanation of the above observation is that, as mentioned in \cite{Exculsive Region:Martin}, the higher-order statistics of the point process formed by the active STs have a marginal effect on the computed Laplace transform of the aggregate interference from all active STs to the typical PR at the origin. Second, the simulated coverage probability of the primary network under the PTA protocol falls between the upper and lower bounds derived in {Theorem~\ref{Theorem 3}} as expected. Third, the PRA protocol outperforms the PTA protocol on the coverage performance of the primary network. Intuitively, this is because that the PRA protocol protects the PRs more directly than the PTA protocol.

\begin{figure*}[!h]
\centerline{\subfigure[$\lambda_0 = 0.01$]{\includegraphics[width
=3.5in]{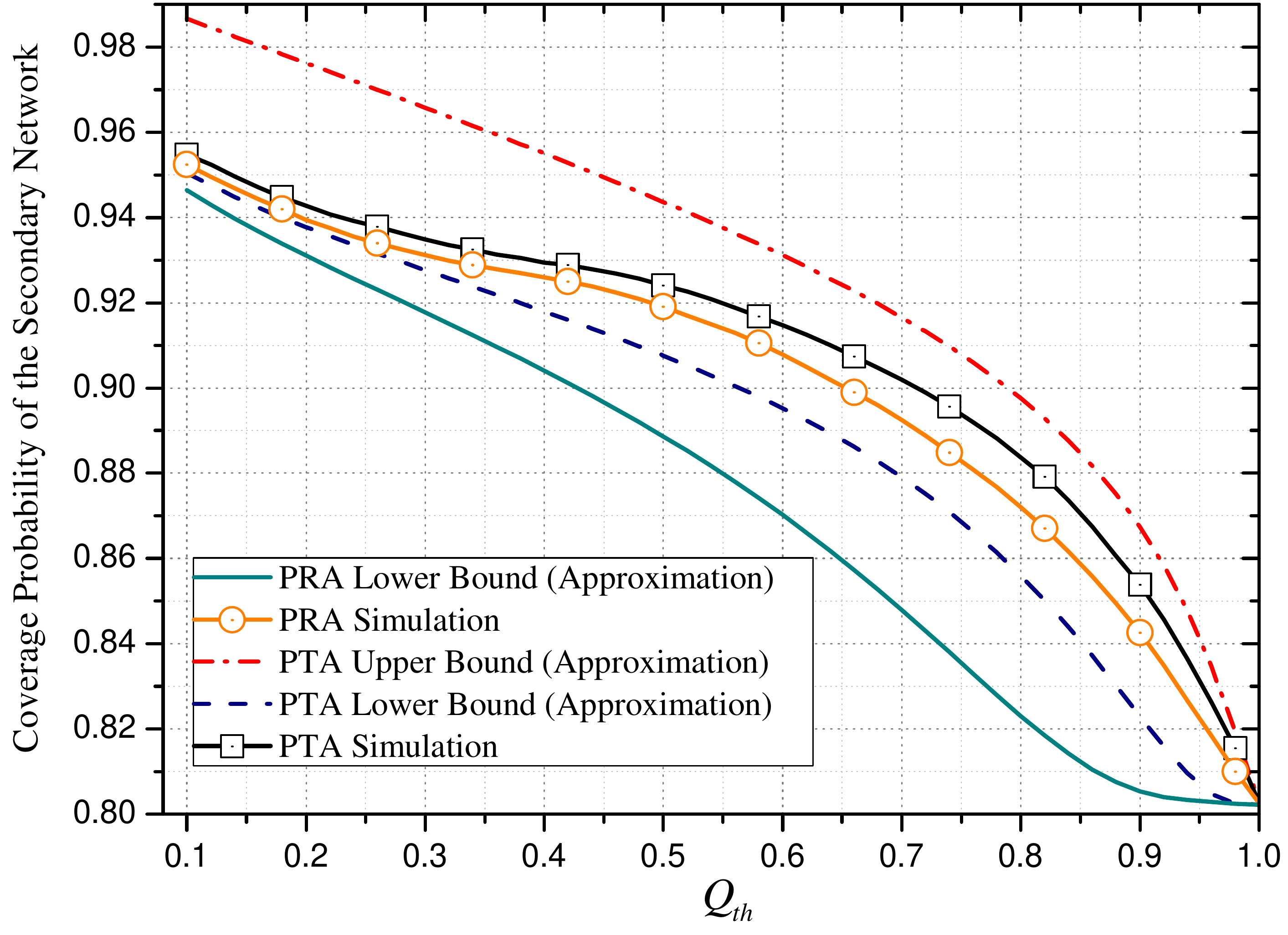}}
\hfil
\subfigure[${\lambda_0 = 0.1}$]{\includegraphics[width=3.45in]{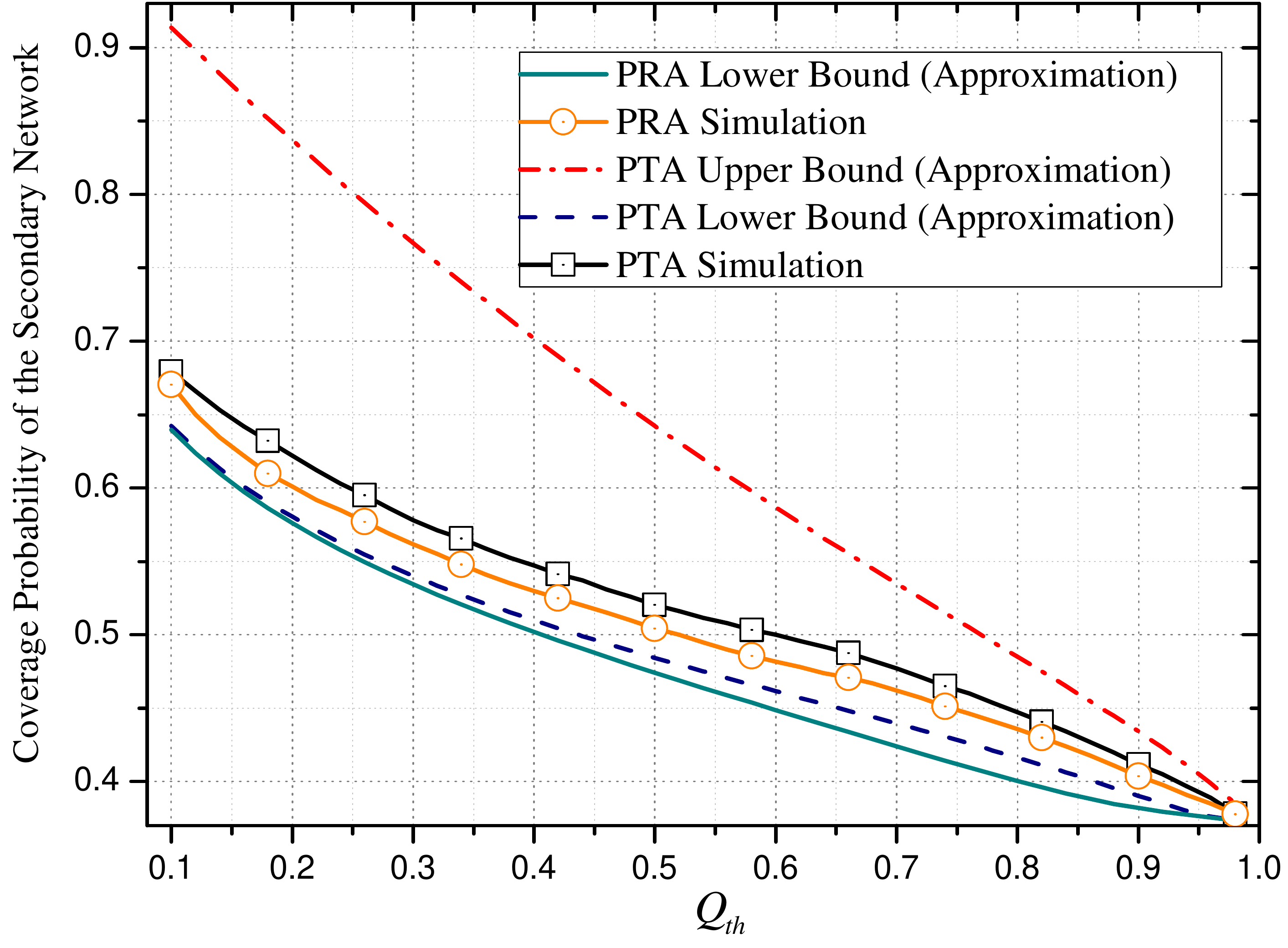}}}
\caption{Coverage probability of the secondary network versus spatial opportunity $Q_{th}$ under the PRA/PTA protocol, with ${Q_{ra} = Q_{ta} = Q_{th}}$.}
\label{Secondary Coverage Probability versus N_th}
\end{figure*}
%

Fig.~\ref{Secondary Coverage Probability versus N_th} compares the analytical results on the coverage probability of the secondary network with the corresponding simulated values under the PRA and PTA protocols, respectively. As observed from Fig.~\ref{Secondary Coverage Probability versus N_th}, the lower and/or upper bounds on the coverage probability of the secondary network derived in {Theorems~\ref{Theorem 4}} and {\ref{Theorem 5}} under the PRA/PTA protocol are effective. It is also observed that the PTA protocol outperforms the PRA protocol on the coverage performance of the secondary network, which is the opposite to the case of the coverage performance of the primary network as shown in Fig.~\ref{Primary Coverage Probability versus N_th}. Intuitively, this is because from the perspective of secondary transmissions, the PTA protocol is more desirable than the PRA protocol since the resulting active STs (and thereby their corresponding SRs at a small distance of $d_s$) are better protected from the active PTs (rather than PRs in the case of the PRA protocol) under the PTA protocol. {Furthermore, we can observe from Fig.~\ref{Secondary Coverage Probability versus N_th} that the coverage probability of the secondary network under the PTA protocol is close to the lower bound in the regime of small ${Q_{th}}$. Intuitively, when ${Q_{th}}$ is small, as discussed in Remark 5.3, the coverage probability of the secondary network under the PTA protocol is dominated by the secondary transmissions around the typical SR at the origin. As such, due to the fact that the density of active STs around the typical SR is close to $\lambda_{s}^{ta}\beta_{ta}$ when ${Q_{th}}$ is sufficiently small, the corresponding lower bound is tight. It is also worth noting that, as can be observed in Fig.~\ref{Secondary Coverage Probability versus N_th}, both the upper and lower bounds converge at ${Q_{th}} = 1$, which is intuitively expected from Theorem \ref{Theorem 5}.}


\begin{figure*}[!h]
\centerline{\subfigure[$\lambda_0 = 0.01$]{\includegraphics[width
=3.5in]{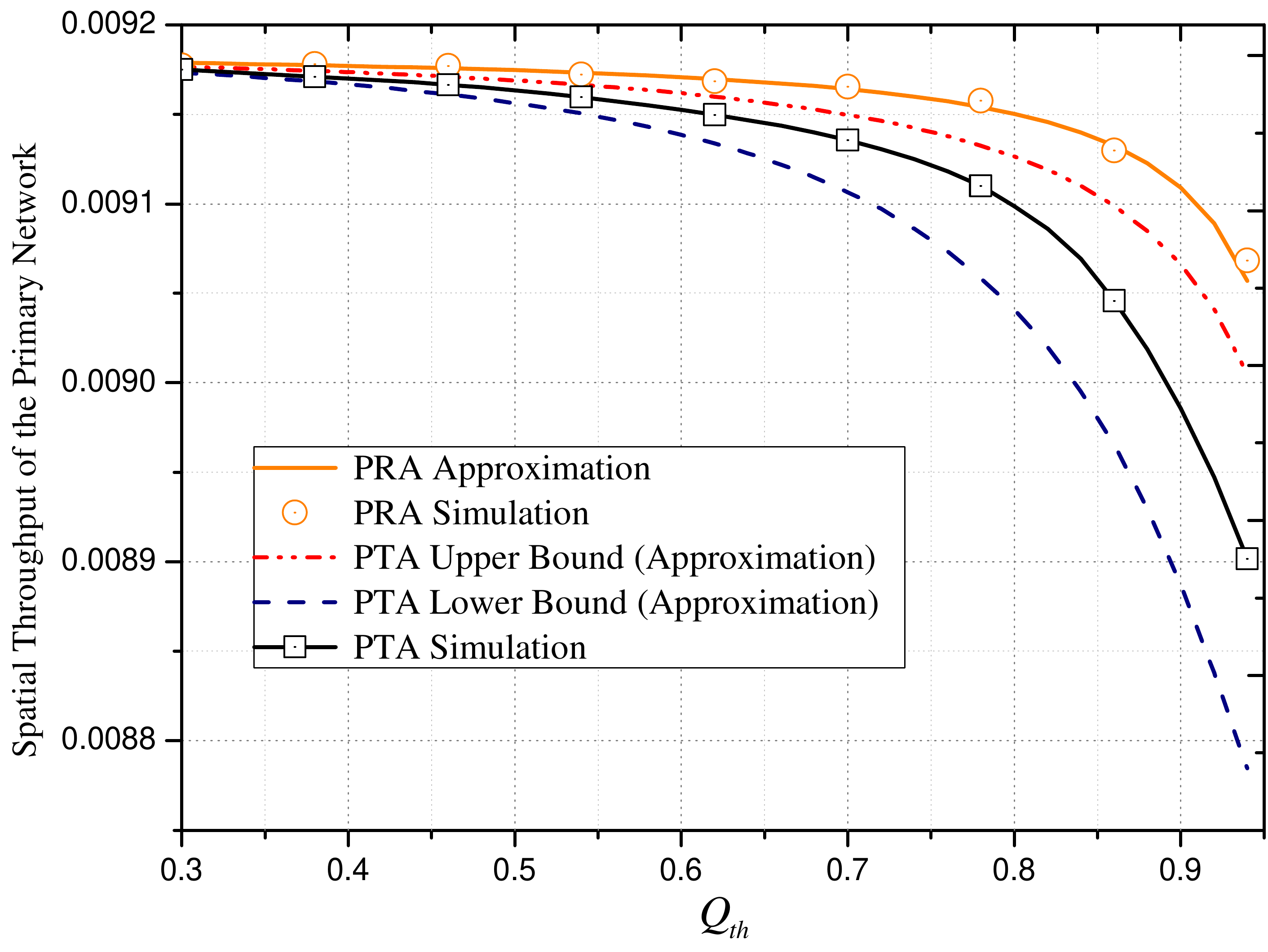}}
\hfil
\subfigure[${\lambda_0 = 0.1}$]{\includegraphics[width=3.5in]{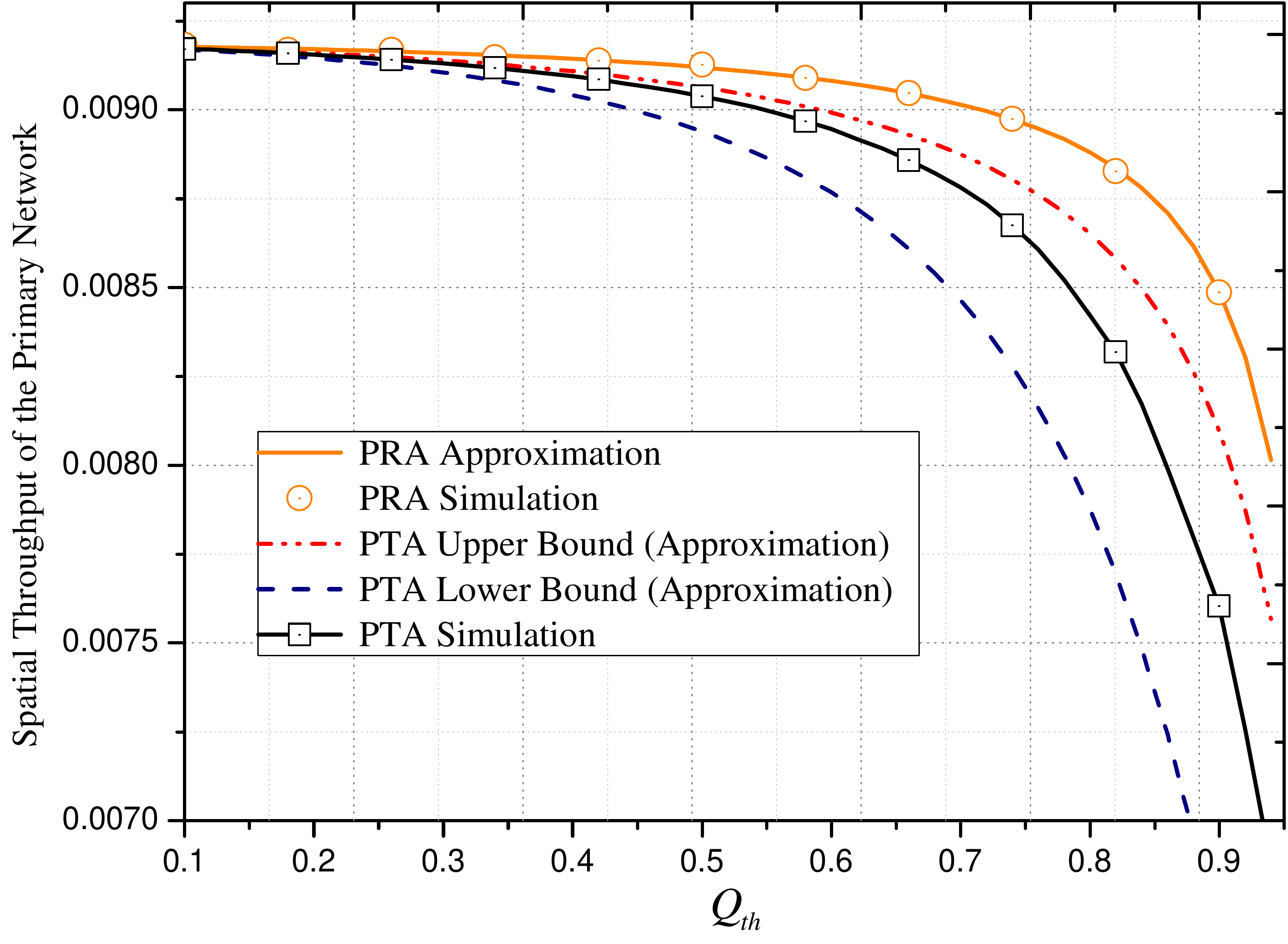}}}
\caption{Spatial throughput of the primary network versus spatial opportunity $Q_{th}$ under the PRA/PTA protocol, with $Q_{ra} = Q_{ta} = Q_{th}$.}
\label{Primary Throughput versus Q}
\end{figure*}

\begin{figure*}[!h]
\centerline{\subfigure[$\lambda_0 = 0.01$]{\includegraphics[width
=3.5in]{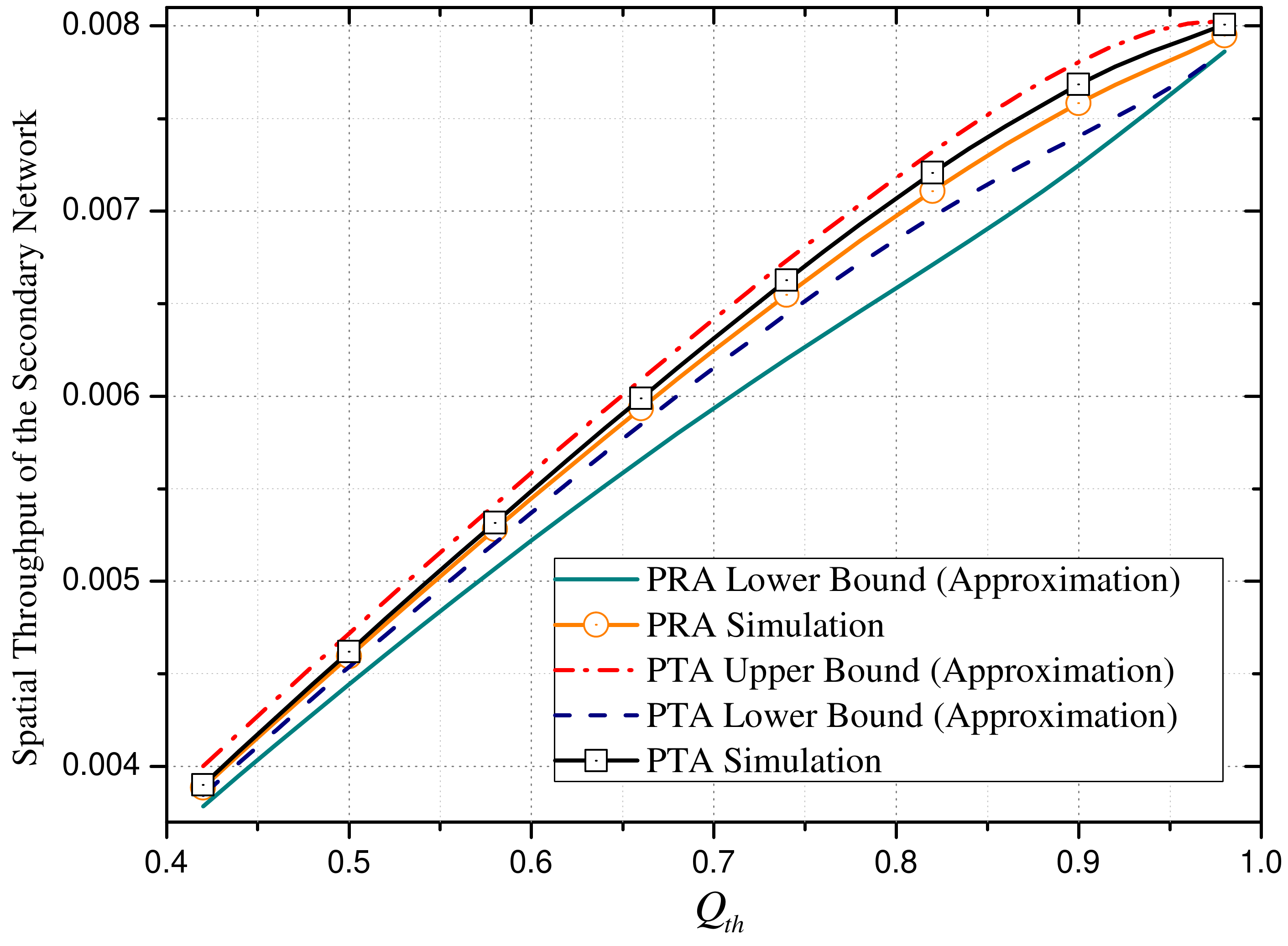}}
\hfil
\subfigure[${\lambda_0 = 0.1}$]{\includegraphics[width=3.5in]{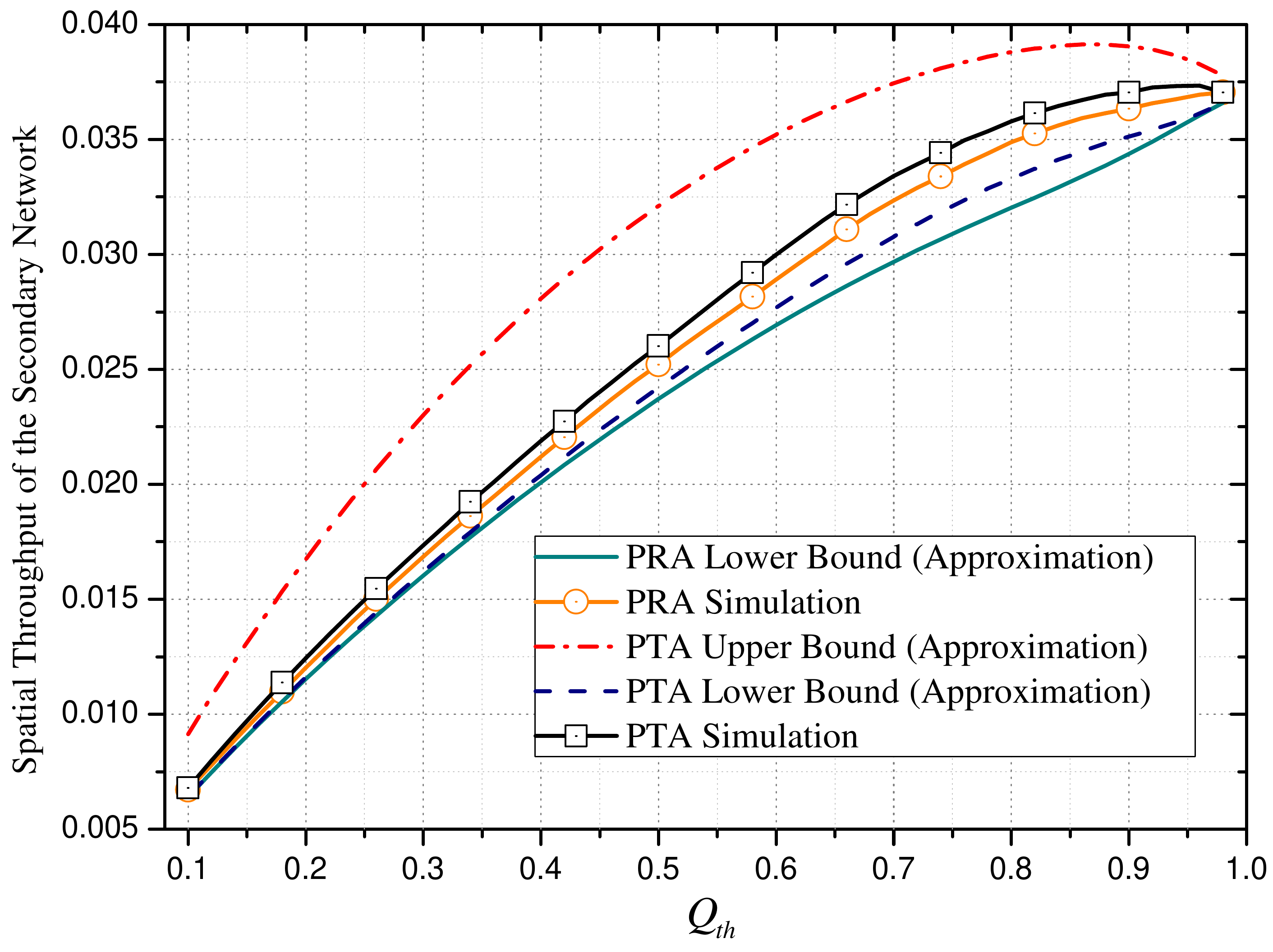}}}
\caption{Spatial throughput of the secondary network versus spatial opportunity $Q_{th}$ under the PRA/PTA protocol, with $Q_{ra} = Q_{ta} = Q_{th}$.}
\label{Secondary Throughput versus Q}
\end{figure*}

\subsection{Spatial Throughput}
Figs. \ref{Primary Throughput versus Q} and \ref{Secondary Throughput versus Q} show the spatial throughput of the primary and secondary networks, respectively, versus spatial opportunity $Q_{th}$ under the PRA/PTA protocol, where we set $Q_{ra} = Q_{ta} = Q_{th}$. Similar discussions for Figs.~\ref{Primary Coverage Probability versus N_th} and \ref{Secondary Coverage Probability versus N_th} can be made for Figs.~\ref{Primary Throughput versus Q} and \ref{Secondary Throughput versus Q}, respectively.
\subsection{Performance Comparison with Exclusion Region Based OSA}
At last, we compare the spectrum-sharing performance of the proposed PRA and PTA protocols with the protocols based on the exclusion regions around PRs (namely ERR) and PTs (namely ERT) proposed in \cite{Exculsive Region:Martin} in terms of primary and secondary network spatial throughput trade-off. Specifically, in a CR network with the ERR or ERT protocol, the STs are allowed to transmit only if they are outside all the exclusion regions of the active PRs or PTs. As such, essentially, the threshold-based OSA protocols can be regarded as ``soft'' versions of the exclusion region based OSA protocols since they take both the distance-dependent attenuation and channel fading effects into account for the activation of STs. Let $D$ denote the radius of the exclusion region around each active PR or PT in the ERR or ERT protocol. Then, according to \cite{Exculsive Region:Martin}, the spatial opportunity for the STs under the ERR or ERT protocol is given by $Q_{err} = Q_{ert} = \exp\left\{- \mu_p \pi D^2\right\}$.
\begin{figure*}[!h]
\centerline{\subfigure[$\lambda_0 = 0.01$]{\includegraphics[width
=3.5in]{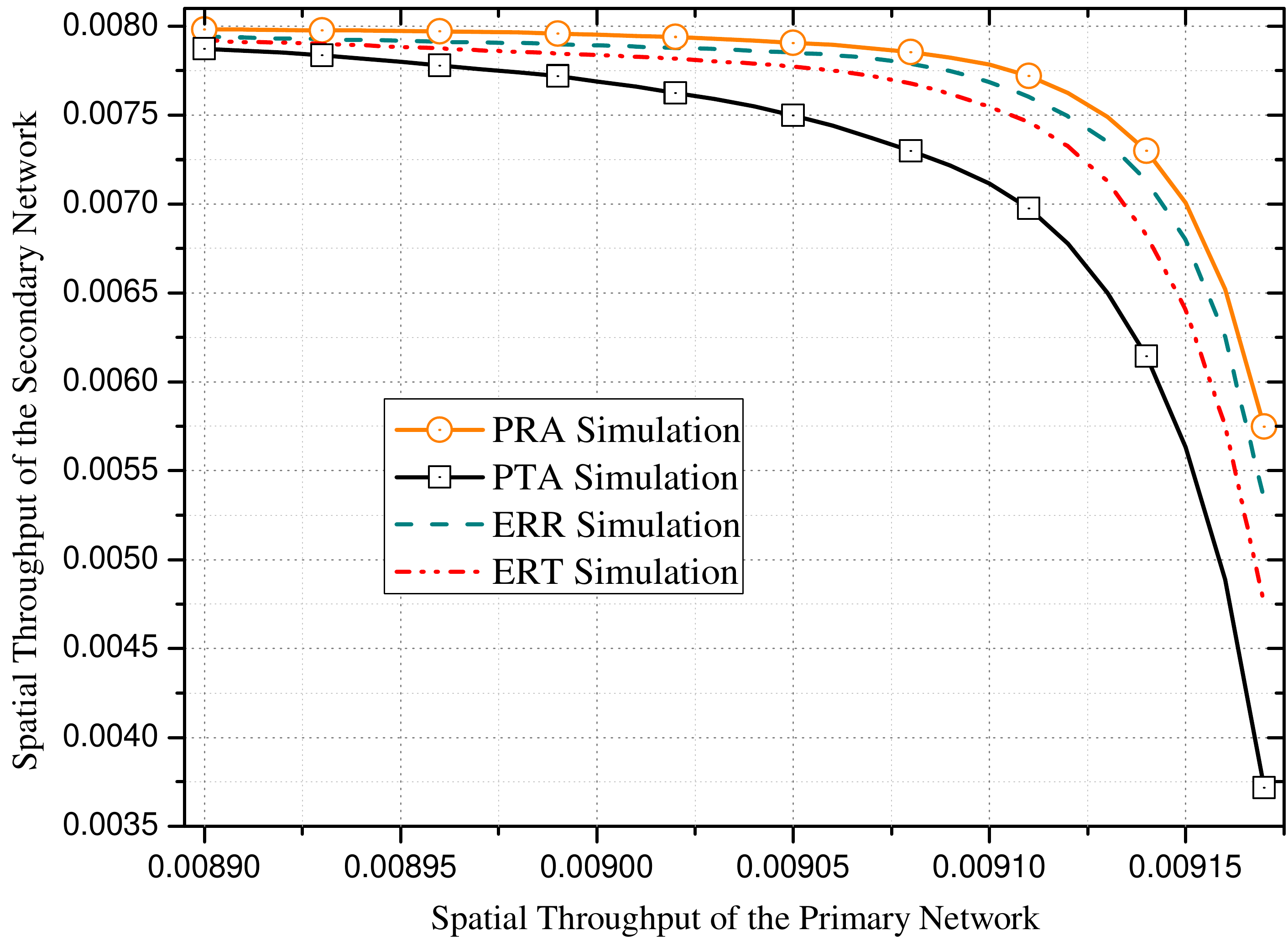}}
\hfil
\subfigure[${\lambda_0 = 0.1}$]{\includegraphics[width=3.5in]{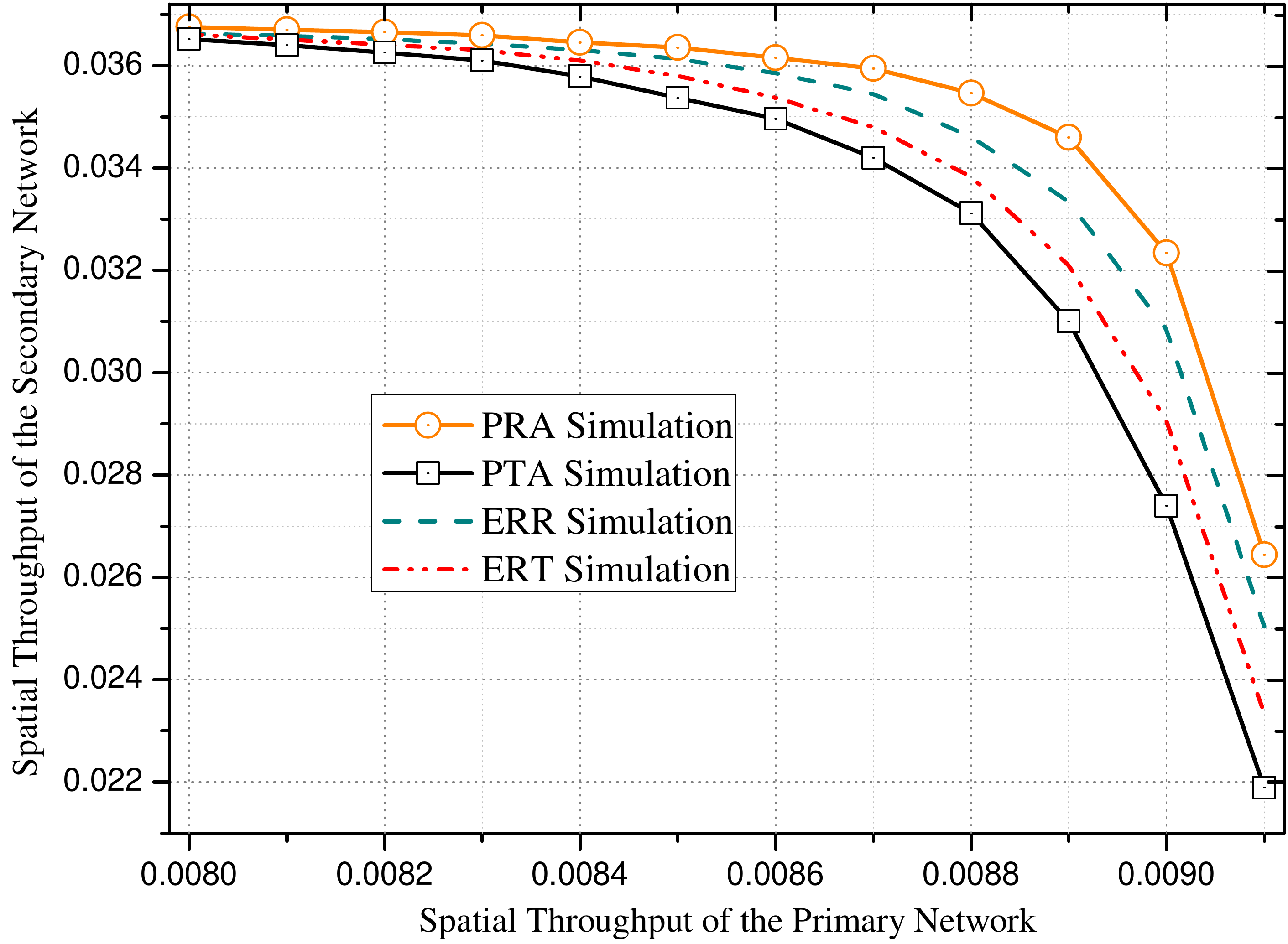}}}
\caption{Simulated spatial throughput trade-off curves for the coexisting primary and secondary networks under the PRA, PTA, ERR and ERT protocols.} \label{Secondary throughput versus Cp}
\end{figure*}
%

Fig.~\ref{Secondary throughput versus Cp} shows
the simulated spatial throughput trade-off curves for the coexisting primary and secondary networks, under the threshold-based OSA protocols and exclusion region based OSA protocols, respectively. As observed from Fig.~\ref{Secondary throughput versus Cp}, the primary versus secondary network spatial throughput trade-off of the PRA/ERT protocol outperforms that of the ERR/PTA protocol. An implication of the above observation is as follows: the threshold-based OSA is more beneficial if the STs are able to detect the active PRs, while the exclusion region based OSA is more favourable if the spatial spectrum holes of the primary network are detected based on the active PTs. {Intuitively, the former is due to the fact that the ``soft'' protection of PRs by PRA protocol based on channel gain from PR is more effective than the ``hard'' protection counterpart by ERR protocol based on exclusive region centered at PR; whereas the latter is because the ERT protocol based on exclusive region centered at PT more effectively protects the corresponding PR as compared to the PTA protocol based on channel gain from PT which may overlook the case when PR is in fact close to ST (but non-detectable due to the faded channel from PT to ST).}

\section{Conclusion}
This paper has studied the performance of spatial OSA in a large-scale overlay CR network. Two threshold-based OSA protocols, namely PRA and PTA, are investigated. By applying tools from stochastic geometry, the spatial opportunity for the secondary network under the PRA/PTA protocol is derived. The conditional distributions of active STs and/or PTs given a typical PR/SR at the origin are then characterized. Based on such results, the coverage probabilities as well as the spatial throughputs of the primary and secondary networks under each of the two proposed protocols are analyzed. It is hoped that the results in this paper will provide new insights to the optimal design of practical overlay based CR networks employing threshold-based OSA.
\appendices
\section{Proof of {Theorem~\ref{Theorem 1}}}\label{appendix 1}
\begin{IEEEproof}
Consider first the case of PRA protocol. Let $\Pi_p^r$ be the set of all active PRs. Then, under the PRA protocol, at an arbitrary location ${\bf{x}} \in \mathbb{R}^2$, the received beacon power $S_{ra}^i ({\bf{x}})$ from the $i$-th active PR with $i \in \Pi_p^r$ is given by
\begin{equation}\label{Received Beacon Signal Strength}
S_{ra}^i ({\bf{x}})= \frac{P_p h_{ra}^i}{|{\bf{X}}_i - {\bf{x}}|^{\alpha}},
\end{equation}
where ${\bf{X}}_i$ is the coordinate of the $i$-th active PR, $h_{ra}^i$ is the power coefficient of the fading channel between ${\bf{X}}_i$ and location $\bf{x}$, and $|{\bf{X}}_i - {\bf{x}}|$ is the corresponding distance. Let $M_{ra}({\bf{x}})$ denote the maximum received beacon power at position $\bf{x}$ as
\begin{equation}\label{Maximum Beacon Signal Strength}
M_{ra}({\bf{x}}) = \max\limits_{i \in \Pi_p^r}S_{ra}^i ({\bf{x}}).
\end{equation}
Then, we derive the spatial opportunity $Q_{ra}$ under the PRA protocol as follows:
\begin{equation}\label{Proof Spatial Opportunity for RA}
\begin{split}
\!\!\!\!\!Q_{ra} & = \Pr\left\{ M_{ra}({\bf{x}}) \leq N_{ra} \right\}\\
& \buildrel{{(a)}}\over= \mathbb{E}\Bigg[1_{\left\{M_{ra}{(\bf{x})} \leq N_{ra}\right\}}\Bigg] \\
& = \mathbb{E}\Bigg[\prod\limits_{i \in \Pi_p^{r}}1_{\left\{S_{ra}^i{(\bf{x})} \leq N_{ra}\right\}}\Bigg]\\
& = \mathbb{E}_{\bf{X}}\left[\prod\limits_{i \in \Pi_p^r}\mathbb{E}_{h}\left[1_{\left\{S_{ra}^i{(\bf{x})} \leq N_{ra}\right\}}\right]\right]\\
& \buildrel{(b)}\over= \exp\left\{- 2 \pi \mu_p \int_0^{\infty}\!\!\left(1 - \Pr\left\{{ h_{ra}^i} \leq \frac{N_{ra} r^{\alpha}}{P_p}\right\}\right)r \textrm{d} r\right\} \\
& = \exp\left\{- 2 \pi \mu_p \int_0^{\infty} e^{-\frac{N_{ra} r^{\alpha}}{P_p}}r \textrm{d}r\right\} \\
& = \exp\left\{- 2 \pi \mu_p \frac{\Gamma(\frac{2}{\alpha})(\frac{P_p}{N_{ra}})^{\frac{2}{\alpha}}}{\alpha} \right\},
\end{split}
\end{equation}
where {$(a)$ follows from the definition of indicator function over random variable,} and $(b)$ follows from the probability generating functional (PGFL) of the PPP defined in \cite{Book: Martin}. By using the same approach as for the case of PRA protocol, $Q_{ta}$ for the PTA protocol can be similiarly obtained. This thus completes the proof of {Theorem~\ref{Theorem 1}}.~
\end{IEEEproof}
\section{Proof of {Theorem~\ref{Theorem 2}}}\label{appendix 2}
\begin{IEEEproof}
With the PRA protocol, for a typical PR at the origin, the SIR is given by
\begin{equation}\label{SIR for PR with RA}
\mathrm{SIR}_p = \frac{P_p h_0 d_p^{- \alpha}}{\sum\limits_{i \in \Pi_p^t}P_p h_i |{\bf{X}}_i|^{-\alpha} + \sum\limits_{j \in \Pi_{s}^{ra}}P_s g_j|{\bf{Y}}_j|^{-\alpha}},
\end{equation}
where $\Pi_p^t$ denotes the set of all active PTs, $\Pi_{s}^{ra}$ denotes the set of all active STs, $h_0$ is the fading channel power coefficient of the typical primary link, $h_i$ is the power coefficient of the fading channel from the $i$-th active PT to the typical PR with $i \in \Pi_p^t$, $g_j$ is the power coefficient of the fading channel from the $j$-th active ST to the typical PR with $j \in \Pi_{s}^{ra}$, ${\bf{X}}_i$ is the coordinate of the $i$-th active PT, and ${\bf{Y}}_j$ is the coordinate of the $j$-th active ST. According to the PRA protocol, at the typical PR, the interference introduced by the $j$-th active ST is constrained as $P_p g_j |{\bf{Y}}_j|^{- \alpha}\leq {N_{ra}}$.
Thus, under Assumption 1, the coverage probability of the primary network with the PRA protocol is given by
\begin{equation}\label{Coverage Probability for Primary with RA}
\begin{split}
\!\!\!\!\!\!\!\tau_p^{ra} & = \Pr\left\{{\mathrm{SIR}}_p \geq \theta_p \Bigg| g_j |{\bf{Y}}_j|^{-\alpha} \leq \frac{N_{ra} }{P_p}\right\}\\
& = \Pr\left\{\frac{P_p h_0 d_p^{- \alpha}}{\sum\limits_{i \in \Pi_p^t}P_p h_i |{\bf{X}}_i|^{-\alpha} + \sum\limits_{j \in \Pi_{s}^{ra}}P_s g_j|{\bf{Y}}_j|^{-\alpha}} \geq \theta_p \Bigg| g_j |{\bf{Y}}_j|^{-\alpha} \leq \frac{N_{ra} }{P_p}\right\}\\
& \buildrel{(a)}\over = {\mathbb{E}}_{\bf{X}}\left[\prod\limits_{i \in \Pi_p^t} {\mathbb{E}}_h\left[e^{-\frac{\theta_p h_i |{\bf{X}}_i|^{-\alpha}}{d_p^{- \alpha}}}\right]\right] \times {\mathbb{E}}_{\bf{Y}} \left[\prod\limits_{j \in \Pi_{s}^{ra}} {\mathbb{E}}_g \left[e^{-\frac{\theta_p P_s g_j |{\bf{Y}}_j|^{-\alpha}}{P_p d_p^{- \alpha}}}\Bigg| g_j \leq \frac{N_{ra} |{\bf{Y}}_j|^{\alpha}}{P_p}\right]\right]\\
& \buildrel{(b)}\over = \exp\left\{-  \frac{2 \pi^2}{\alpha \sin\left(\frac{2 \pi}{\alpha}\right)}  \mu_p\theta_p^{\frac{2}{\alpha}} d_p^2 \right\}\\
& \quad\times \exp\left\{- 2 \pi \int_0^{\infty}\left(1 - \int_0^{\frac{N_{ra} u^{\alpha}}{P_p}} e^{- \frac{\theta_p P_s g u^{- \alpha}}{P_p d_p^{- \alpha}}} \times \frac{e^{-g}}{1 - e^{- {\frac{N_{ra} u^{\alpha}}{P_p}}}}dg \right) \lambda_{ra}^{\mathbf{R}}(u) u \textrm{d}u\right\}\\
& = \exp\left\{-  \frac{2 \pi^2}{\alpha \sin\left(\frac{2 \pi}{\alpha}\right)}  \theta_p^{\frac{2}{\alpha}} d_p^2 \left(\mu_p + \lambda_s^{ra} \left(\frac{P_s}{P_p}\right)^{\frac{2}{\alpha}}\right) \right\} \times \exp\left\{\frac{2 \pi}{\alpha} \lambda_s^{ra} \left(\frac{P_p}{N_{ra}}\right)^{\frac{2}{\alpha}}\Gamma(\frac{2}{\alpha})\right\}\\
&\quad\times \exp\left\{ - 2 \pi \lambda_s^{ra} \int_0^{\infty} \frac{\frac{P_p u^{\alpha}}{\theta_p P_s d_p^{\alpha}}}{1 + \frac{P_p u^{\alpha}}{\theta_p P_s d_p^{\alpha}}} \times
\frac{ e^{- \frac{\theta_p P_s N_{ra} d_p^{\alpha}}{P_p^2}}}{e^{\frac{N_{ra} u^{\alpha}}{P_p}}}u \textrm{d}u\right\},
\end{split}
\end{equation}
where $(a)$ follows from Assumption 1 that the point processes formed by the active PTs and STs are assumed to be independent, and $(b)$ follows from the fact that the probability density function of $g$ conditioned on $g \leq t$ is given by
\[f(g|g \leq t) = \frac{e^{-g}}{1 - e^{-t}}.\]
This thus completes the proof of {Theorem~\ref{Theorem 2}}.
\end{IEEEproof}

\section{Proof of Lemma \ref{Lemma 3}}\label{appendix 3}
\begin{IEEEproof}
To prove Lemma \ref{Lemma 3}, we define
\begin{equation}\label{psuedo interference}
G_j({\bf{x}}) = \frac{1}{1 + \frac{P_p |{\bf{Y}}_j - {\bf{x}}|^{\alpha}}{\theta_p P_s d_p^{\alpha}}},
\end{equation}
as the pseudo interference perceived at an arbitrary location ${\bf{x}} \in \mathbb{R}^2$ introduced by the $j$-th active ST at location ${\bf{Y}}_j$ with $j \in \Pi_s^{ta}$, where $\Pi_s^{ta}$ denotes the set of all active STs under the PTA protocol. Let $P_I({\mathbf{R}}_p)$ and $P_I({\mathbf{T}}_p)$ be the aggregate pseudo interference from all active STs perceived at ${\mathbf{R}}_p$ and ${\mathbf{T}}_p$, respectively, under the PTA protocol. Then, we have
\begin{equation}\label{agregate interference at R}
\begin{split}
\mathbb{E}\left[P_I({\mathbf{R}}_p)\right] &= \mathbb{E}\left[\sum\limits_{j \in \Pi_s^{ta}}G_j({\mathbf{R}}_p)\right] \buildrel{(a)}\over= 2 \pi \int_0^{\infty}\frac{\lambda_{ta}^{{\mathbf{R}}_p}(u)}{1 + \frac{P_p u^{\alpha}}{\theta_p P_s d_p^{\alpha}}} u \textrm{d}u,
\end{split}
\end{equation}
and
\begin{equation}\label{agregate interference at T}
\begin{split}
\mathbb{E}\left[P_I({\mathbf{T}}_p)\right] = \mathbb{E}\left[\sum\limits_{j \in \Pi_s^{ta}}G_j({\mathbf{T}}_p)\right] &\buildrel{(b)}\over= 2 \pi \int_0^{\infty}\frac{\lambda_{ta}^{{\mathbf{T}}_p}(u)}{1 + \frac{P_p u^{\alpha}}{\theta_p P_s d_p^{\alpha}}} u \textrm{d}u,
\end{split}
\end{equation}
where $(a)$ and $(b)$ follow from the Campbell's Theorem \cite{Book: Martin}. As a result, based on (\ref{agregate interference at R}) and (\ref{agregate interference at T}), it follows that to prove (\ref{lower Bound on the intensity with TA}), we only need to show that
\begin{equation}\label{PIR geq PIT}
\mathbb{E}\left[P_I({\mathbf{R}}_p)\right]\geq \mathbb{E}\left[P_I({\mathbf{T}}_p)\right].
\end{equation}

To prove (\ref{PIR geq PIT}), we partition the plane $\mathbb{R}^2$ into infinite number of equal-size squares of the same area $\triangle{s}$ and index them as illustrated in Fig.~\ref{Partition}. Without loss of generality, we focus on the $i$-th pair of squares $s_{i}^1$ and $s_{i}^2$ both of which are at a distance\footnote{The distance from $s_{i}^1$ or $s_{i}^2$ to $\mathcal{M}$ refers to that from the center of $s_{i}^1$ or $s_{i}^2$ to $\mathcal{M}$.} of $r$ to $\mathcal{M}$ as illustrated in Fig.~\ref{Interference of TA}, where $\mathcal{M}$ is the perpendicular bisector to the line between ${\mathbf{R}}_p$ and ${\mathbf{T}}_p$. Then, as $\Delta{s}\to 0$, the point processes formed by the active STs in $s_{i}^1$ and $s_{i}^2$ asymptotically follow two PPPs with density $\lambda_{ta}^{{\mathbf{T}}_p}(r_2)$ and $\lambda_{ta}^{{\mathbf{T}}_p}(r_1)$, respectively, where $r_1$ denotes the distance between $s_{i}^2$ and ${\mathbf{T}}_p$, and $r_2$ denotes that between $s_{i}^1$ and ${\mathbf{T}}_p$.\footnote{By symmetry, as illustrated in Fig.~\ref{Interference of TA}, the distances from $s_{i}^1$ and $s_{i}^2$ to ${\mathbf{R}}_p$ are also given by $r_1$ and $r_2$, respectively.}
\begin{figure*}[!h]
\centerline{\subfigure[]{\includegraphics[width
=3.7in]{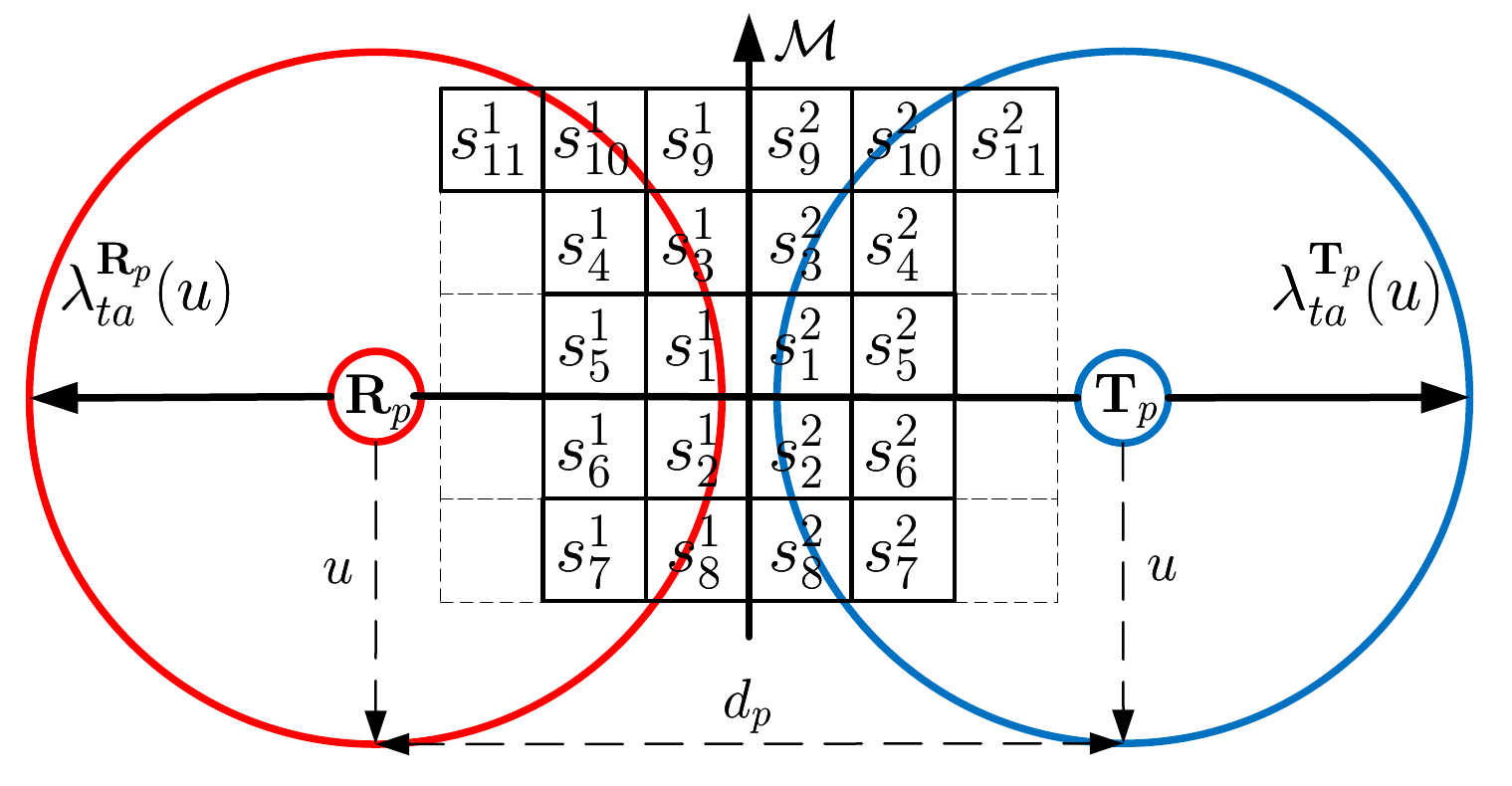}
\label{Partition}}
\hfil
\subfigure[]{\includegraphics[width=3.4in]{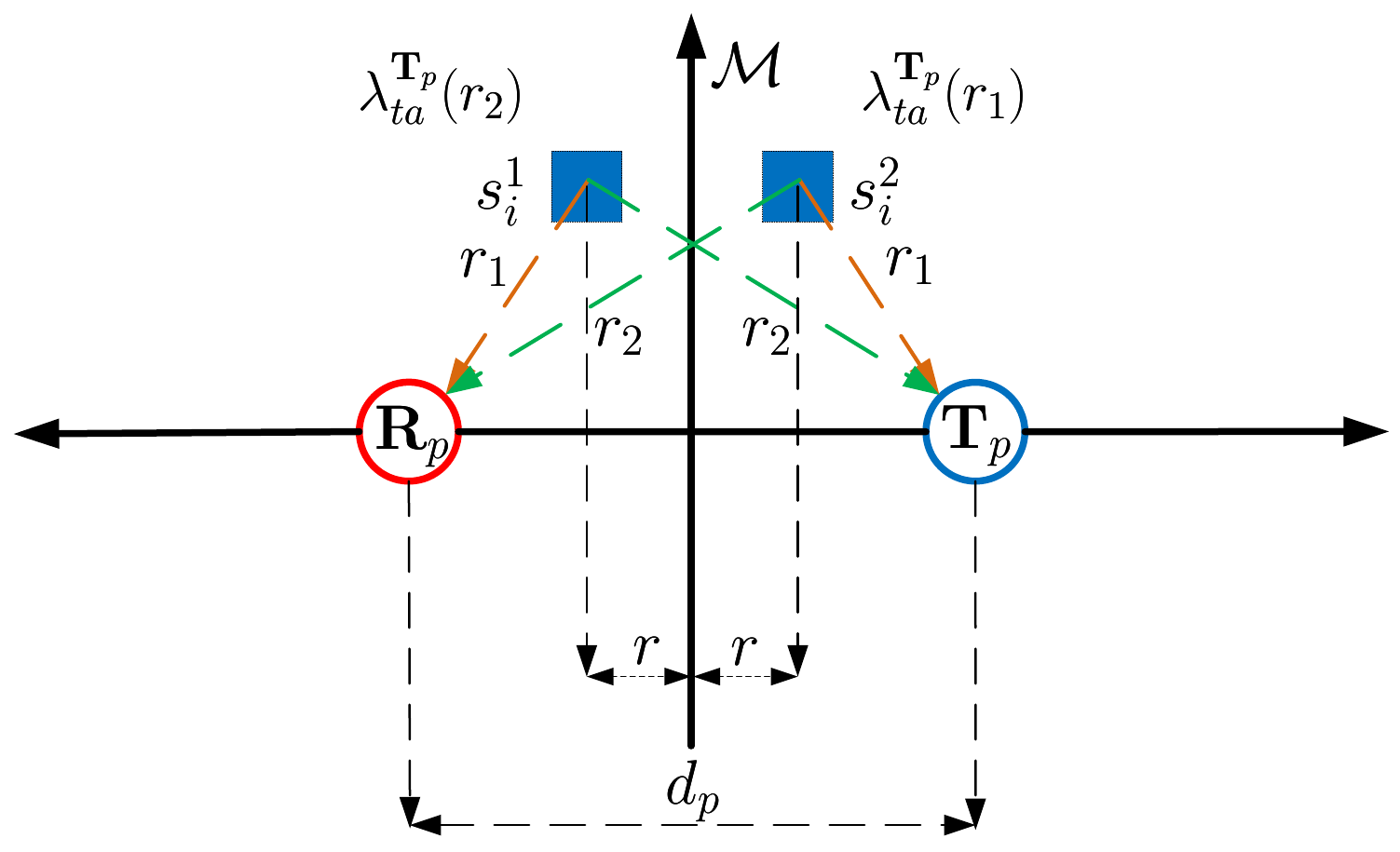}
\label{Interference of TA}}}
\caption{(a) Partition of the plane in $\mathbb{R}^2$; and (b) the aggregate pseudo interference from active STs in the union of $s_{i}^1$ and $s_{i}^2$ perceived at ${\mathbf{R}_p}$ or ${\mathbf{T}_p}$.}
\label{Proof}
\end{figure*}

Let $P_{I(s_{i}^1, s_{i}^2)}({\mathbf{R}}_p)$ be the aggregate pseudo interference perceived at ${\mathbf{R}}_p$ from the active STs in the union of $s_{i}^1$ and $s_{i}^2$. Then, we have
\begin{equation}\label{s1 + s2 interference at R}
\mathop {\lim }\limits_{\Delta s \to 0} \mathbb{E}\left[P_{I(s_{i}^1, s_{i}^2)}({\mathbf{R}}_p)\right]
\buildrel{(a)}\over= \mathop {\lim }\limits_{\Delta s \to 0}\left(\frac{\lambda_{ta}^{{\mathbf{T}}_p}(r_1)}{1 + \frac{P_p r_2^{\alpha}}{\theta_p P_s d_p^{\alpha}}} + \frac{\lambda_{ta}^{{\mathbf{T}}_p}(r_2)}{1 + \frac{P_p r_1^{\alpha}}{\theta_p P_s d_p^{\alpha}}}\right) \times\Delta{s},
\end{equation}
where $(a)$ follows from the Campbell's Theorem.

Similarly, let $P_{I(s_{i}^1, s_{i}^2)}({\mathbf{T}}_p)$ be the aggregate pseudo interference perceived at ${\mathbf{T}}_p$ from the active STs in the union of $s_{i}^1$ and $s_{i}^2$. Then, we have
\begin{equation}\label{s1 + s2 interference at T}
\mathop {\lim }\limits_{\Delta s \to 0} \mathbb{E}\left[P_{I(s_{i}^1, s_{i}^2)}({\mathbf{T}}_p)\right]
= \mathop {\lim }\limits_{\Delta s \to 0}\left(\frac{\lambda_{ta}^{{\mathbf{T}}_p}(r_2)}{1 + \frac{P_p r_2^{\alpha}}{\theta_p P_s d_p^{\alpha}}} + \frac{\lambda_{ta}^{{\mathbf{T}}_p}(r_1)}{1 + \frac{P_p r_1^{\alpha}}{\theta_p P_s d_p^{\alpha}}}\right) \times\Delta{s}.
\end{equation}
With (\ref{s1 + s2 interference at R}) and (\ref{s1 + s2 interference at T}), we obtain that
\begin{equation}\label{subtraction}
\begin{split}
&\quad\mathop {\lim }\limits_{\Delta s \to 0} \left(\mathbb{E}\left[P_{I(s_{i}^1, s_{i}^2)}({\mathbf{R}}_p)\right] - \mathbb{E}\left[P_{I(s_{i}^1, s_{i}^2)}({\mathbf{T}}_p)\right]\right)\\
&=\mathop {\lim }\limits_{\Delta s \to 0}\left(\lambda_{ta}^{{\mathbf{T}}_p}(r_2) - \lambda_{ta}^{{\mathbf{T}}_p}(r_1) \right)\times \left(\frac{1}{1 + \frac{P_p r_1^{\alpha}}{\theta_p P_s d_p^{\alpha}}} - \frac{1}{1 + \frac{P_p r_2^{\alpha}}{\theta_p P_s d_p^{\alpha}}} \right)\times\Delta{s}\\
&\buildrel{(a)}\over\geq 0,
\end{split}
\end{equation}
where $(a)$ follows from the fact that $r_1 \leq r_2$.

Furthermore, it is worth noting that
\begin{equation}\label{IR)}
\mathbb{E}\left[P_{I}({\mathbf{R}}_p)\right] = \mathop {\lim }\limits_{\Delta s \to 0} \sum\limits_{i = 1}^{\infty} {\mathbb{E}\left[P_{I(s_{i}^1, s_{i}^2)}({\mathbf{R}}_p)\right]},
\end{equation}
and
\begin{equation}\label{IT}
\mathbb{E}\left[P_I(\mathbf{T}_p)\right] = \mathop {\lim }\limits_{\Delta s \to 0} \sum\limits_{i = 1}^{\infty} {\mathbb{E}\left[P_{I(s_{i}^1, s_{i}^2)}({\mathbf{T}}_p)\right]}.
\end{equation}
As such, the following inequality holds:
\begin{equation}\label{Final}
\begin{split}
&\quad \mathbb{E}\left[P_{I}({\mathbf{R}}_p)\right] - \mathbb{E}\left[P_{I}({\mathbf{T}}_p)\right]\\
&= \mathop {\lim }\limits_{\Delta s \to 0} \sum\limits_{i = 1}^{\infty} {\mathbb{E}\left[P_{I(s_{i}^1, s_{i}^2)}({\mathbf{R}}_p)\right]} - \mathop {\lim }\limits_{\Delta s \to 0} \sum\limits_{i = 1}^{\infty} {\mathbb{E}\left[P_{I(s_{i}^1, s_{i}^2)}({\mathbf{T}}_p)\right]}\\
& = \mathop {\lim }\limits_{\Delta s \to 0} \sum\limits_{i = 1}^{\infty} {\left(\mathbb{E}\left[P_{I(s_{i}^1, s_{i}^2)}({\mathbf{R}}_p)\right] - \mathbb{E}\left[P_{I(s_{i}^1, s_{i}^2)}({\mathbf{T}}_p)\right]\right)}\\
&\buildrel{(a)}\over\geq 0,
\end{split}
\end{equation}
where $(a)$ follows from (\ref{subtraction}). Therefore, we have $\mathbb{E}\left[P_{I}({\mathbf{R}}_p)\right]\geq\mathbb{E}\left[P_{I}({\mathbf{T}}_p)\right]$, i.e.,
\[
\int_0^{\infty}\frac{\lambda_{ta}^{{\mathbf{R}}_p}(u)}{1 + \frac{P_p u^{\alpha}}{\theta_p P_s d_p^{\alpha}}} u \textrm{d}u \geq \int_0^{\infty}\frac{\lambda_{ta}^{{\mathbf{T}}_p}(u)}{1 + \frac{P_p u^{\alpha}}{\theta_p P_s d_p^{\alpha}}} u \textrm{d}u.
\]
This thus completes the proof of Lemma~\ref{Lemma 3}.
\end{IEEEproof}
\section{Proof of {Theorem~\ref{Theorem 3}}}\label{appendix 4}
\begin{IEEEproof}
With the PTA protocol, for a typical PR at the origin, the SIR is given by
\begin{equation}\label{SIR for PR with TA}
\mathrm{SIR}_p = \frac{P_p h_0 d_p^{- \alpha}}{\sum\limits_{i \in \Pi_p^t}P_p h_i |{\bf{X}}_i|^{-\alpha} + \sum\limits_{j \in \Pi_{s}^{ta}}P_s g_j|{\bf{Y}}_j|^{-\alpha}},
\end{equation}
where $\Pi_p^t$ denotes the set of all active PTs, $\Pi_{s}^{ta}$ denotes the set of all active STs. Let $q_j$ be the power coefficient of the fading channel from the $j$-th active ST to the typical PT with $j \in \Pi_{s}^{ta}$.
Then, according to the PTA protocol, the interference introduced by the $j$-th active ST is constrained at the typical PT as $P_p q_j |{\bf{Y}}_j - {\mathbf{T}_p}|^{- \alpha}\leq {N_{ta}}$. It should be noted that $q_j$ and $g_j$ are statistically independent for any given $j \in \Pi_{s}^{ta}$. Therefore, the interference introduced by the $j$-th active ST can be arbitrary high at the typical PR. As such, under Assumption 1, the coverage probability of the primary network with the PTA protocol is given by
\vspace{-0.2in}
\begin{equation}\label{Coverage Probability for Primary with TA}
\begin{split}
\tau_p^{ta} & = \Pr\left\{{\mathrm{SIR}}_p \geq \theta_p \Bigg| q_j |{\bf{Y}}_j-{\mathbf{T}_p}|^{- \alpha} \leq \frac{N_{ta} }{P_p}\right\}\\
& = \Pr\left\{\frac{P_p h_0 d_p^{- \alpha}}{\sum\limits_{i \in \Pi_p^t}P_p h_i |{\bf{X}}_i|^{-\alpha} + \sum\limits_{j \in \Pi_{s}^{ta}}P_s g_j|{\bf{Y}}_j|^{-\alpha}} \geq \theta_p \right\}\\
& \buildrel{(a)}\over = {\mathbb{E}}_{\bf{X}}\left[\prod\limits_{i \in \Pi_p^t} {\mathbb{E}}_h\left[e^{-\frac{\theta_p h_i |{\bf{X}}_i|^{-\alpha}}{d_p^{- \alpha}}}\right]\right] \times {\mathbb{E}}_{\bf{Y}} \left[\prod\limits_{j \in \Pi_{s}^{ta}} {\mathbb{E}}_g \left[e^{-\frac{\theta_p P_s g_j |{\bf{Y}}_j|^{-\alpha}}{P_p d_p^{- \alpha}}}\right]\right]\\
& = \exp\left\{-  \frac{2 \pi^2}{\alpha \sin\left(\frac{2 \pi}{\alpha}\right)}  \mu_p\theta_p^{\frac{2}{\alpha}} d_p^2 \right\} \times \exp\left\{- 2 \pi \int_0^{\infty}\frac{\lambda_{ta}^{{\mathbf{R}}_p}(u)}{1 + \frac{P_p u^{\alpha}}{\theta_p P_s d_p^{\alpha}}} u \textrm{d}u\right\},
\end{split}
\end{equation}
where $(a)$ follows from Assumption 1 that the active STs are assumed to be distributed independently with the active PTs.

Then, by applying Lemma~\ref{Lemma 2} and \ref{Lemma 3} to (\ref{Coverage Probability for Primary with TA}), we obtain the lower and upper bounds on the coverage probability of the primary network under the PTA protocol as
\begin{equation}\label{Coverage Probability for Primary with TA Lower}
\begin{split}
\tau_p^{ta} & \geq \exp\left\{-  \frac{2 \pi^2}{\alpha \sin\left(\frac{2 \pi}{\alpha}\right)}  \mu_p\theta_p^{\frac{2}{\alpha}} d_p^2 \right\}\times \exp\left\{- 2 \pi \int_0^{\infty}\frac{\lambda_s^{ta}\left(1 - e^{ - \frac{N_{ta} (u + d_p)^{\alpha}}{P_p}}\right)}{1 + \frac{P_p u^{\alpha}}{\theta_p P_s d_p^{\alpha}}} u du\right\}\\
& = \exp\left\{-  \frac{2 \pi^2}{\alpha \sin\left(\frac{2 \pi}{\alpha}\right)}  \theta_p^{\frac{2}{\alpha}} d_p^2 \left(\mu_p + \lambda_s^{ta} \left(\frac{P_s}{P_p}\right)^{\frac{2}{\alpha}}\right) \right\}\\ 
&\qquad\qquad\qquad\qquad\times \exp\left\{ 2 \pi \lambda_s^{ta} \int_0^{\infty} \left(\frac{e^{ - \frac{N_{ta} (u + d_p)^{\alpha}}{P_p}}}{1 + \frac{P_p u^{\alpha}}{\theta_p P_s d_p^{\alpha}}} \right)u \textrm{d}u\right\},
\end{split}
\end{equation}
and
\begin{equation}\label{Coverage Probability for Primary with TA Upper}
\begin{split}
\tau_p^{ta} & \leq \exp\left\{-  \frac{2 \pi^2}{\alpha \sin\left(\frac{2 \pi}{\alpha}\right)}  \mu_p\theta_p^{\frac{2}{\alpha}} d_p^2 \right\} \times \exp\left\{- 2 \pi \int_0^{\infty}\frac{\lambda_{ta}^{{\mathbf{T}}_p}(u)}{1 + \frac{P_p u^{\alpha}}{\theta_p P_s d_p^{\alpha}}} u du\right\}\\
& = \exp\left\{-  \frac{2 \pi^2}{\alpha \sin\left(\frac{2 \pi}{\alpha}\right)}  \theta_p^{\frac{2}{\alpha}} d_p^2 \left(\mu_p + \lambda_s^{ta} \left(\frac{P_s}{P_p}\right)^{\frac{2}{\alpha}}\right) \right\}\\
&\qquad\qquad\qquad\qquad\times \exp\left\{ 2 \pi \lambda_s^{ta} \int_0^{\infty} \left(\frac{e^{ - \frac{N_{ta} u^{\alpha}}{P_p}}}{1 + \frac{P_p u^{\alpha}}{\theta_p P_s d_p^{\alpha}}} \right)u \textrm{d}u\right\},
\end{split}
\end{equation}
respectively.
This thus completes the proof of {Theorem~\ref{Theorem 3}}.
\end{IEEEproof}
\section{Proof of Lemma \ref{Lemma 5}}\label{appendix 5}
\begin{IEEEproof}
We first consider the case of PRA protocol. Under the PRA protocol, according to Lemma~\ref{Lemma 4}, ${\Psi}_{ra}^{{\mathbf{T}}_s}(r)$ follows a HPPP and thus is isotropic around ${\mathbf{T}}_s$. Then, by the isotropy of ${\Psi}_{ra}^{{\mathbf{T}}_s}(r)$, it can be easily verified that ${\Phi}_{ra}^{{\mathbf{T}}_s}(r)$ follows a point process that is also isotropic around~${\mathbf{T}_s}$.

To prove the lower and upper bounds on ${\lambda^{{\mathbf{T}}_s}_{ra}}({r})$, by the isotropy of ${\Phi}_{ra}^{{\mathbf{T}}_s}(r)$, we consider the spatial opportunity ${Q'_{ra}}({\bf{x}}_s)$ of the STs at an arbitrary location ${\bf{x}}_s$, where $|{\bf{x}}_s-{\mathbf{T}_s}| = r$. Let ${\psi}_{ra}^{{\bf{x}}_s}(t)$ be the average density of the PPP formed by the active PRs on a circle of radius $t$ centered at ${\bf{x}}_s$ as illustrated in Fig.~\ref{Proof of Lemma 5}. Then, similar to the proof of {Theorem~\ref{Theorem 1}}, we obtain a lower bound on ${Q'_{ra}}({\bf{x}}_s)$ as
\begin{equation}\label{Q'ra lower bound}
\begin{split}
Q'_{ra}({\bf{x}}_s)& = \exp\left\{- 2 \pi \int_0^{\infty} e^{-\frac{N_{ra} t^{\alpha}}{P_p}} {\psi}_{ra}^{{\bf{x}}_s}(t) t \textrm{d}t\right\} \\
& \buildrel{(a)}\over \geq \exp\left\{- 2 \pi \mu_p \int_0^{\infty} e^{-\frac{N_{ra} t^{\alpha}}{P_p}} t \textrm{d}t\right\} \\
& = Q_{ra},
\end{split}
\end{equation}
where $(a)$ follows from Lemma \ref{Lemma 4} that ${\psi}_{ra}^{{\bf{x}}_s}(t) \leq \mu_p$.
\begin{figure}[!h]
\centering
\includegraphics[width=2.7in]{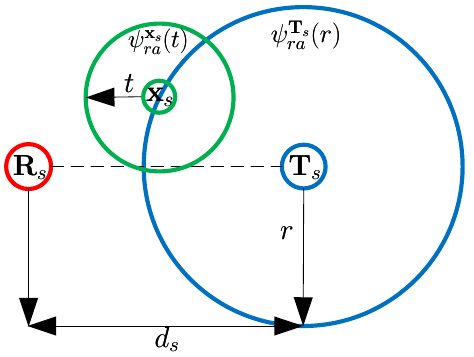}
\caption{Proof of the lower and upper bounds on $Q'_{ra}({\bf{x}}_s)$.} \label{Proof of Lemma 5}
\end{figure}

To prove the upper bound on $Q'_{ra}({\bf{x}}_s)$, we define
\begin{equation}\label{pseudo interference}
G_j({\bf{x}}) = e^{-\frac{N_{ra} |{\bf{X}}_j - {\bf{x}}|^{\alpha}}{P_p}}
\end{equation}
as the pseudo interference perceived at an arbitrary location ${\bf{x}} \in \mathbb{R}^2$ introduced by the $j$-th active PR at location ${\bf{X}}_j$ with $j \in \Pi_{p}^r$, where $\Pi_{p}^r$ denotes the set of all active PRs. Then, by applying a similar proof as for Lemma \ref{Lemma 3}, the following inequality is obtained:
\begin{equation}\label{Q'ra inequality}
\int_0^{\infty} e^{-\frac{N_{ra} t^{\alpha}}{P_p}} {\psi}_{ra}^{{\bf{x}}_s}(t) t \textrm{d}t \geq \int_0^{\infty} e^{-\frac{N_{ra} r^{\alpha}}{P_p}} {\psi}_{ra}^{{\mathbf{T}}_s}({r}) r \textrm{d}r.
\end{equation}
As such, based on (\ref{Q'ra inequality}), we derive the upper bound on $Q'_{ra}({\bf{x}}_s)$ as
\begin{equation}\label{Q'ra Ts}
\begin{split}
Q'_{ra}({\bf{x}}_s)& = \exp\left\{- 2 \pi \int_0^{\infty} e^{-\frac{N_{ra} t^{\alpha}}{P_p}} {\psi}_{ra}^{{\bf{x}}_s}(t) t \textrm{d}t\right\} \\
& \leq \exp\left\{- 2 \pi \int_0^{\infty} e^{-\frac{N_{ra} r^{\alpha}}{P_p}} {\psi}_{ra}^{{\mathbf{T}}_s}({r}) r \textrm{d}r\right\} \\
& = Q_{ra}\beta_{ra}.
\end{split}
\end{equation}
Finally, with (\ref{Q'ra lower bound}) and (\ref{Q'ra Ts}), we obtain that
\[
\lambda_s^{ra} \leq {\lambda^{{\mathbf{T}}_s}_{ra}}({r}) \leq \lambda_s^{ra}\beta_{ra}.
\]
This thus proves the lower and upper bounds on $\lambda^{{\mathbf{T}}_s}_{ra}({r})$ under the PRA protocol.

For the case of PTA protocol, with a similar proof as the above for the PRA protocol, (\ref{Secondary Coverage Secondary Density with TA}) can be obtained. This thus completes the proof of Lemma \ref{Lemma 5}.
\end{IEEEproof}


\section*{Acknowledgment}
The authors would like to thank the anonymous reviewers for their helpful comments.

The first author would like to thank Heng Su, Feng Jiang and Jie Chen at University of California, Irvine for their valuable suggestions. The first author would also like to thank Professor Syed Ali Jafar at University of California, Irvine for his guidance and training on information theory and interference alignment.



%

\end{document}